\documentclass[aps,prl,reprint,superscriptaddress]{revtex4-2}

\usepackage{graphicx}
\usepackage{textcomp}
\newcommand{\qty}[2]{#1\,#2}

\begin{document}
\title{High-Q Magnetic Levitation and Control of Superconducting Microspheres at Millikelvin Temperatures}

\author{J. Hofer}
\email[]{joachim.hofer@univie.ac.at}
\affiliation{Faculty of Physics, Vienna Center for Quantum Science and Technology (VCQ), University of Vienna, A-1090 Vienna, Austria}
\affiliation{Institute for Quantum Optics and Quantum Information (IQOQI), Austrian Academy of Sciences, A-1090 Vienna, Austria}
\author{R. Gross}
\affiliation{Walther-Meißner-Institut, Bayerische Akademie der Wissenschaften, D-85748 Garching, Germany}
\affiliation{Physik-Department, Technische Universität München, D-85748 Garching, Germany}
\affiliation{Munich Center for Quantum Science and Technology (MCQST), D-80799 München, Germany}
\author{G. Higgins}
\affiliation{Institute for Quantum Optics and Quantum Information (IQOQI), Austrian Academy of Sciences, A-1090 Vienna, Austria}
\affiliation{Department of Microtechnology and Nanoscience (MC2), Chalmers University of Technology, S-412 96 Gothenburg, Sweden}
\author{H. Huebl}
\affiliation{Walther-Meißner-Institut, Bayerische Akademie der Wissenschaften, D-85748 Garching, Germany}
\affiliation{Physik-Department, Technische Universität München, D-85748 Garching, Germany}
\affiliation{Munich Center for Quantum Science and Technology (MCQST), D-80799 München, Germany}
\author{O. F. Kieler}
\affiliation{Physikalisch-Technische Bundesanstalt (PTB), D-38116 Braunschweig, Germany}
\author{R. Kleiner}
\affiliation{Physikalisches Institut, Center for Quantum Science (CQ) and LISA\textsuperscript{+}, University of Tuebingen, D-72076 Tuebingen, Germany}
\author{D. Koelle}
\affiliation{Physikalisches Institut, Center for Quantum Science (CQ) and LISA\textsuperscript{+}, University of Tuebingen, D-72076 Tuebingen, Germany}
\author{P. Schmidt}
\affiliation{Institute for Quantum Optics and Quantum Information (IQOQI), Austrian Academy of Sciences, A-1090 Vienna, Austria}
\author{J. A. Slater}
\altaffiliation{Currently at: QuTech, Delft University of Technology, Delft, The Netherlands}
\altaffiliation{Faculty of Physics, Vienna Center for Quantum Science and Technology (VCQ), University of Vienna, A-1090 Vienna, Austria}
\author{M. Trupke}
\affiliation{Faculty of Physics, Vienna Center for Quantum Science and Technology (VCQ), University of Vienna, A-1090 Vienna, Austria}
\author{K. Uhl}
\affiliation{Physikalisches Institut, Center for Quantum Science (CQ) and LISA\textsuperscript{+}, University of Tuebingen, D-72076 Tuebingen, Germany}
\author{T. Weimann}
\affiliation{Physikalisch-Technische Bundesanstalt (PTB), D-38116 Braunschweig, Germany}
\author{W. Wieczorek}
\affiliation{Faculty of Physics, Vienna Center for Quantum Science and Technology (VCQ), University of Vienna, A-1090 Vienna, Austria}
\affiliation{Department of Microtechnology and Nanoscience (MC2), Chalmers University of Technology, S-412 96 Gothenburg, Sweden}
\author{M. Aspelmeyer}
\affiliation{Faculty of Physics, Vienna Center for Quantum Science and Technology (VCQ), University of Vienna, A-1090 Vienna, Austria}
\affiliation{Institute for Quantum Optics and Quantum Information (IQOQI), Austrian Academy of Sciences, A-1090 Vienna, Austria}

\date{\today}

\begin{abstract}
We report the levitation of a superconducting lead-tin sphere with \qty{100}{\textmu m} diameter (corresponding to a mass of \qty{5.6}{\textmu g}) in a static magnetic trap formed by two coils in an anti-Helmholtz configuration, with adjustable resonance frequencies up to \qty{240}{Hz}. The center-of-mass motion of the sphere is monitored magnetically using a dc superconducting quantum interference device as well as optically and exhibits quality factors of up to $2.6\times 10^{7}$. We also demonstrate 3D magnetic feedback control of the motion of the sphere. The setup is housed in a dilution refrigerator operating at \qty{15}{mK}. By implementing a cryogenic vibration isolation system, we can attenuate environmental vibrations at \qty{200}{Hz} by approximately 7 orders of magnitude. The combination of low temperature, large mass and high quality factor provides a promising platform for testing quantum physics in previously unexplored regimes with high mass and long coherence times. 
\end{abstract}

\maketitle

Diamagnets and superconductors partially expel magnetic fields, allowing stable levitation in field minima \cite{Brandt1989}. A prominent application of magnetically levitated systems is the use as ultraprecise acceleration sensors, most notably in the superconducting gravimeter \cite{Goodkind1999}, which relies on the levitation of centimeter-sized hollow superconducting spheres in a stable magnetic field generated by superconducting coils in persistent current mode. More recently, several proposals \cite{Isart2012,Cirio2012} have highlighted the potential of magnetic levitation for tests of quantum physics with macroscopic, micrometer-sized objects. 

Successfully preparing a magnetically levitated particle in a quantum state requires a magnetic trap with low damping, low heating rates and the ability to control the mechanical motion. Some of these features are already present in recent demonstrations of magnetically levitated systems, such as the levitation of permanent magnets above a superconducting surface \cite{Vinante2020, Gieseler2020} and the levitation of diamagnets in the field of permanent magnets at cryogenic temperatures \cite{Leng2021} or room temperature \cite{Urso2021}. However, as of yet no system has combined these features. Furthermore, as typical levitation frequencies for these systems range from subhertz to kilohertz, the heating rate is dominated by environmental vibrations.

One of the most promising avenues toward the quantum regime is the levitation of a type-I (or zero-field-cooled type-II) superconductor in a magnetic field produced by persistent currents, as this avoids dissipation due to hysteresis and eddy currents, which is inherent to levitation schemes involving ferromagnets or flux-pinned type-II superconductors \cite{Wang2019, Brandt1988}. Using persistent currents should result in an extremely stable trap, as both the drift and noise can be kept extremely low \cite{Goodkind1999, Van1999, Britton2016}. Working at millikelvin temperatures also naturally enables coupling to nonlinear quantum systems such as superconducting qubits \cite{Isart2012}, and miniaturizing the trap architecture \cite{Latorre2020, Latorre2022, Latorre2022_2} might lead to fully on-chip coupled quantum systems \cite{Pino2018}.
In comparison with optical levitation, which has already been established as a means for studying massive objects in the quantum regime \cite{Delic2020,Magrini2021,Tebbenjohanns2021}, magnetostatic levitation has the potential to access a new parameter regime of even larger mass \cite{Hofer2019,Navau2021} and longer coherence times \cite{Isart2012,Pino2018}: While the size of optically levitated particles is, in practice, limited by the available laser power to the micrometer scale \cite{Moore2017}, magnetic levitation can support train-scale objects \cite{Moon1995} and due to the static nature of the fields there is no heating from photon absorption.

\begin{figure}
\includegraphics[width=\columnwidth]{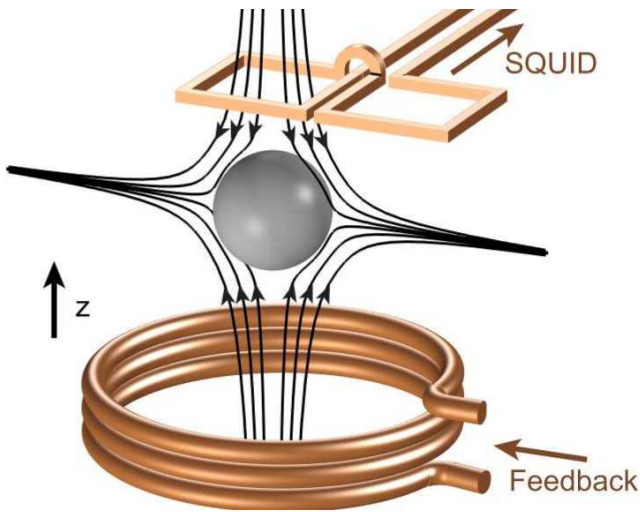}
\caption{\label{fig1} Conceptual representation of the experiment showing the levitated sphere, the pickup coil and the feedback coil. The trap field is depicted by its field lines.}
\end{figure}

A first step toward a fully superconducting platform was presented in \cite{Waarde2016}, where a lead particle was levitated in the magnetic field of a superconducting coil. This approach, however, was limited in performance by the presence of cryogenic exchange gas at \qty{4}{K} and environmental vibrations.
Here, we report an experiment that provides orders of magnitude improvements in both damping and vibration isolation. We demonstrate the levitation of a superconducting sphere in the magnetic field generated by a superconducting coil. 
We detect the center of mass (c.m.) motion of the particle magnetically by inductive coupling to a superconducting quantum interference device (SQUID) via a pickup coil (cf.\ Fig.~\ref{fig1}) and demonstrate feedback control of all three c.m. translational degrees of freedom by real-time processing of the SQUID signal.
We find that the motion of the particle has a low damping rate, corresponding to quality factors of up to $\mathrm{2.6\times 10^{7}}$. We also implement a custom vibration isolation system to reduce heating from environmental vibrations. Finally, we discuss the limitations of our current setup and the improvements that are necessary for ground state cooling.

\paragraph{Trapping.---}
The energy of an object of volume $V$ with permeability $\mu$ in an applied magnetic field $B_{0}$ in free space can be approximated by $-1/2V(\mu-\mu_{0})B_{0}^{2}$ \cite{Brandt1989}, where $\mu_0$ is the permeability of free space.
Because $\nabla^2B_{0}^{2}\geq 0$, stable levitation is possible in a field minimum for $\mu<\mu_{0}$, i.e. for diamagnets.
A superconducting sphere can act as a perfect diamagnet ($\mu=0$), because it reacts to a change in the applied magnetic field by forming screening currents on its surface, counteracting the applied field and keeping the interior of the sphere field free.

For a sphere trapped in the field of an anti-Helmholtz coil, the resulting potential close to the field minimum is harmonic and two of the c.m. modes are degenerate \cite{Isart2012,Hofer2019}; to lift the degeneracy we use elliptical coils \cite{Supp}. The $z$ direction is along the coil axis, which coincides with the vertical direction, while $x$ and $y$ are perpendicular to $z$, with $x$ along the major axis of the coils. We use numerical simulations \cite{Supp} to model the magnetic field created by such an arrangement and confirm that the magnetic field near the center between the coils is well described by $(b_{x} x,b_{y} y, b_{z} z)$, where the field gradients $b_i$ are proportional to the trap current with $|b_{x}|<|b_{y}|<|b_{z}|$ and $b_{z} = - (b_{x} + b_{y})$. This field shape constitutes a three-dimensional harmonic trap for a levitated superconducting sphere with the c.m. frequencies \cite{Supp}
\begin{equation}
\label{eq1}
f_i = \sqrt{\frac{3}{8\pi^2\mu_0\rho}}|b_i|,
\end{equation}
where $\rho$ is the density of the sphere. 

Our setup is housed in a dilution refrigerator (Bluefors BF-LD400) with a loaded base temperature of approximately \qty{15}{mK}. 
The microspheres are commercial (EasySpheres) solder balls with diameter \qty{100(6)}{\textmu m}, made of a 90-10 lead-tin alloy with density $\rho=\qty{10.9\times 10^{3}}{\mathrm{kg/m^{3}}}$. Lead-tin is a type-II superconductor with a critical temperature of approximately \qty{7}{K} \cite{Livingston1963}. 
A single sphere is initially placed in a 3D-printed polylactide bowl glued to the lower trap coil, such that it rests around \qty{1}{mm} below the trap center. The bowl helps to prevent accidental loss of the sphere during assembly and between measurement runs. After cooling down the system to millikelvin temperatures, lift-off is achieved by increasing the coil current until the upward magnetic force on the sphere is stronger than the downward gravitational and adhesive forces. 
The trap coils are surrounded by an aluminum shield with small openings for optical access and wires, which screens the sphere from magnetic field fluctuations when the aluminum is superconducting. To suppress field fluctuations from the trapping coils and feedback coil, we use a low pass filter and attenuation stages, respectively \cite{Supp}. 

\paragraph{Readout.---}
A planar Niobium thin film gradiometric pickup coil consisting of two square loops positioned approximately \qty{400}{\textmu m} above the trap center is used to magnetically probe the particle motion. Each loop is \qty{145}{\textmu m}\,$\times$\,\qty{145}{\textmu m} and the separation between the loops is \qty{3}{\textmu m}. As the particle moves, it induces a current in the pickup loop \cite{Supp}, which is connected with Nb wires to the input coil of a SQUID current sensor \cite{Storm2020}. The SQUID has an intrinsic flux noise level of $S_\Phi^{1/2} = $ \qty{0.8}{\textmu$\mathrm{\Phi_{0}/\sqrt{Hz}}$}, where $\Phi_{0}\approx\qty{2.1\times 10^{-15}}{\mathrm{Wb}}$ is the flux quantum. When the SQUID is incorporated into the setup and the pickup loop is connected, the noise rises to $S_\Phi^{1/2} \approx$ \qty{10}{\textmu$\mathrm{\Phi_{0}/\sqrt{Hz}}$}. We believe this is due to external field fluctuations inductively coupling into the wires connecting the pickup loop and the SQUID. With a commercial SQUID (Supracon CSblue), that was used for some of the measurements, the noise floor in the setup further increased by an order of magnitude. 

To calibrate the SQUID signal, we optically record the motion of the sphere by illuminating it and imaging its shadow on a camera, while simultaneously recording the motion with the SQUID \cite{Supp}. The coupling strengths (flux induced in the SQUID per sphere displacement) for the different motional modes depend on the position of the pickup loop in relation to the sphere as well as the trap gradients, i.e. mechanical frequencies.
The highest measured coupling strength was approximately \qty{$13\times 10^{3}$}{$\mathrm{\Phi_{0}/m}$} for the $z$ mode and \qty{$3\times 10^{3}$}{$\mathrm{\Phi_{0}/m}$} for the $x$ and $y$ modes. Together with a flux noise floor of \qty{10}{\textmu$\mathrm{\Phi_{0}/\sqrt{Hz}}$} these coupling strengths correspond to a displacement sensitivity of approximately \qty{1}{$\mathrm{nm/\sqrt{Hz}}$} for the $z$ mode.
During these measurements the separation between pickup loop and sphere was approximately \qty{400}{\textmu m}. We estimate that, with a smaller separation and an optimized pickup loop geometry, an improvement of 4 orders of magnitude is possible \cite{Supp}.

\begin{figure}
\includegraphics[width=\columnwidth]{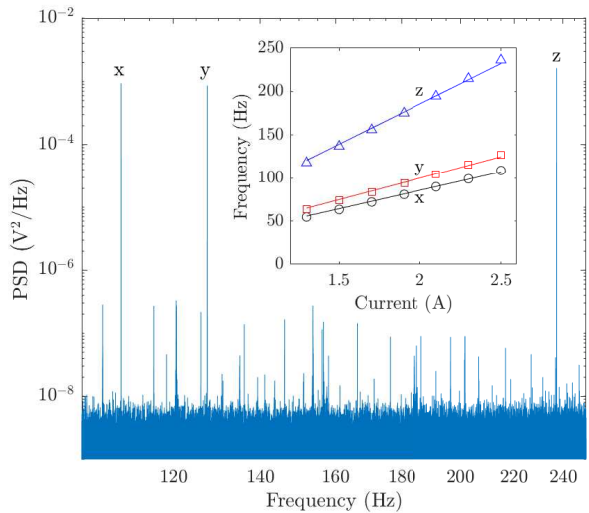}
\caption{\label{fig2} The PSD of the SQUID signal displays three peaks which correspond to the c.m. modes of the particle. The motion of the particle is excited to approximately one micrometer rms displacement in this figure - measurement noise prevents us from resolving the equilibrium motion of the particle. The inset shows the linear dependence of the c.m. frequencies on the trap current. The solid lines are zero-intercept linear fits.}
\end{figure}

\paragraph{Mechanical frequencies.---}
The c.m. frequencies of the particle are visible in the power spectral density (PSD) of the SQUID signal, as shown in Fig.~\ref{fig2}. The c.m. frequencies depend linearly on the magnetic field gradient of the trap [see Eq.~(\ref{eq1})] and, thus, the current in the trap coils, as shown in the inset.
The highest frequency reached in the present setup is \qty{240}{Hz} (corresponding to $b_{z}\approx$ \qty{150}{T/m}), limited by the maximum current output of the low-noise current source. We can trap at lower frequencies, down to $f_z\approx$ \qty{20}{Hz}, where the lower limit stems from the gravity-induced shift of the trapping position and our trap geometry. However, at low trapping frequencies, the sphere is strongly excited by environmental vibrations.

\begin{figure}
\includegraphics[width=\columnwidth]{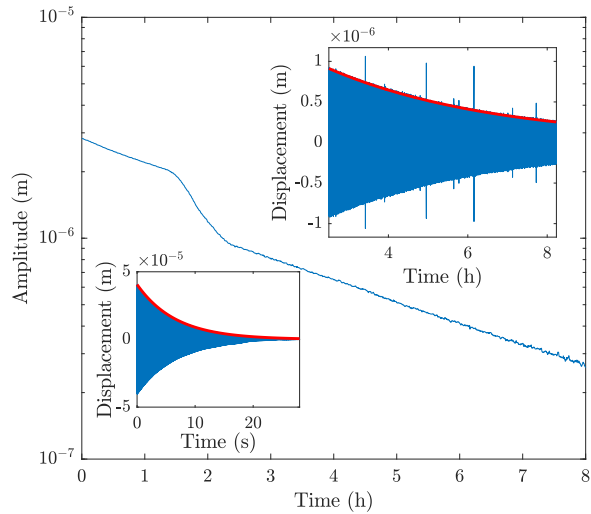}
\caption{\label{fig3} Ringdown measurement of the $z$ mode. The upper right inset shows a linear plot for the latter part of the data, starting at t=\qty{2.5}{h}. The red line is a fit to an exponential decay. The spikes correspond to a brief loss of the SQUID lock, they were removed for the amplitude plot. The lower left inset shows a ringdown while direct feedback is applied (note that the $x$ axis scale is in seconds).}
\end{figure}

We can apply a magnetic feedback force on the levitated sphere by processing the SQUID signal on an FPGA (STEMlab 125-14) and applying a feedback current to a small coil with approximately 20 windings positioned approximately \qty{1}{mm} below the trap center \cite{Supp}. As measurement noise prevents us from resolving the equilibrium c.m. motion of the particle, we use feedback here only as a means to quickly adjust the amplitudes and prepare other measurements (cf. Fig.~\ref{fig3}, inset).

\paragraph{Quality factor.---}
We measure the quality factor by performing ringdown measurements, where the initial starting amplitude is set by applying an appropriate feedback signal before the start of the measurement. 
We do see occasional jumps in the damping rate (cf.\ Fig.~\ref{fig3}), which we attribute to the detaching and attaching of flux lines to a pinning center. Unpinned flux lines can move within the superconductor (flux creep), a mechanism that is known to cause damping in the levitation of flux-pinned superconductors or magnets \cite{Brandt1988, Nemoshkalenko1990}. Although we do not apply a magnetic field during the cooldown, there is a nonzero background field, and hence, we expect some frozen-in flux to be present in the levitated sphere. 
As a consequence, the measured quality factors vary with time and between measurement runs, but they are generally between $1\times 10^{7}$ and $2.5\times 10^{7}$ for the $z$ mode and about half as high for the $x$ and $y$ modes. The highest quality factor we have measured is $2.6\times 10^{7}$ at \qty{212}{Hz}, corresponding to a dissipation rate of \qty{$5\times 10^{-5}$}{$\mathrm{s^{-1}}$}, and likely still limited by flux creep inside the particle.
We also consider other possible contributions to the damping \cite{Supp}, but find that the expected contributions cannot explain the measured values. 
In future experiments we plan to use a particle made of a type-I superconductor such as monocrystalline lead, which will expel all magnetic flux via the Meissner effect and prevent flux creep.

\paragraph{Sensing.---}
A massive system acting as a harmonic oscillator can be used for force and acceleration sensing, with the sensitivities depending on the mass $m$ and the dissipation rate $\gamma$ of the oscillator. In equilibrium with a thermal bath at temperature $T_{0}$, the thermal force noise is given by $\sqrt{S_{FF}^\mathrm{th}}=\sqrt{4k_{B}T_{0}m\gamma}$ \cite{Kubo1966}, where $k_{B}$ is Boltzmann's constant. The dissipation rate $\gamma$ is related to the quality factor $Q$ and the angular mechanical resonance frequency $\omega_0$ by $\gamma=\omega_0/Q$. The corresponding acceleration sensitivity is $\sqrt{S_{FF}^\mathrm{th}}/m$. For a particle of mass $m\approx\qty{5.6}{$\textmu g}, assuming it is in thermal equilibrium with its surroundings in the dilution refrigerator at \qty{15}{mK}, this sets the limits for force and acceleration sensing at \qty{$5\times 10^{-19}$}{$\mathrm{N/\sqrt{Hz}}$} and \qty{$9\times 10^{-12}$}{$\mathrm{g/\sqrt{Hz}}$}, using our highest measured quality factor $Q=2.6\times 10^{7}$ at \qty{212}{Hz}. 
Our current experimental sensitivities are approximately 1 order of magnitude above the thermal noise, limited by drifts in the trap current and measurement noise \cite{Supp}.

\begin{figure}
\includegraphics[width=\columnwidth]{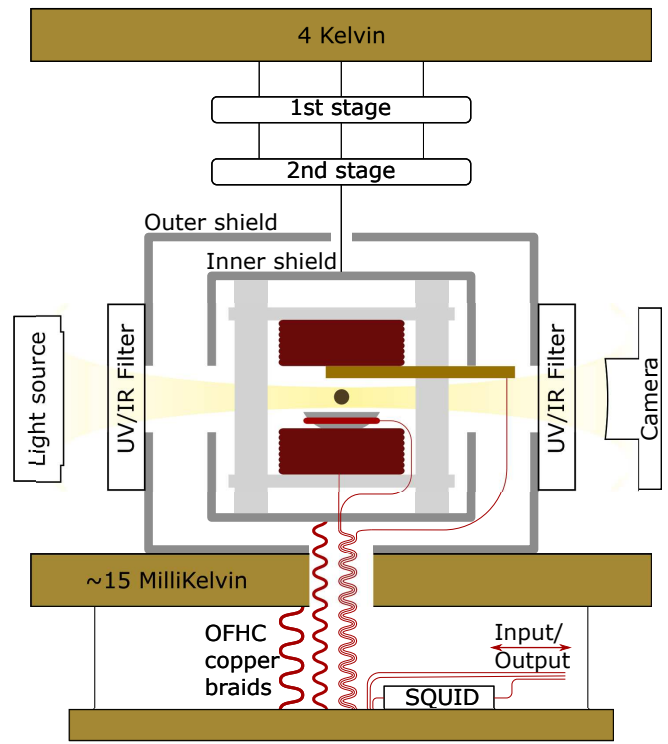}
\caption{\label{fig4} Sketch of the setup including the vibration isolation system (not to scale). The light source is turned on only for calibration measurements. The UV-IR cutoff filters on the millikelvin stage prevent ambient radiation from heating the levitated sphere. Thermal connections are made with oxygen free high conductivity (OFHC) copper braids.}
\end{figure}

\paragraph{Cryogenic vibration isolation.---}
Vibrations of the cryostat accelerate the trapping coils and, thus, the trap center, effecting a force onto the levitated sphere. Denoting the vibrational PSD by $S_{\epsilon\epsilon}$, the displacement PSD of our sphere becomes $|\chi(\omega)|^2m^2\omega_0^4S_{\epsilon\epsilon}(\omega)$,
where $\chi$ denotes the mechanical susceptibility.
From independent accelerometer measurements on the cryostat, we estimate $\sqrt{S_{\epsilon\epsilon}}\approx$ \qty{$1\times 10^{-10}$}{$\mathrm{m/\sqrt{Hz}}$} at \qty{200}{Hz}, which would result in a peak height of \qty{$2.6\times 10^{-3}$}{$\mathrm{m/\sqrt{Hz}}$} and a root-mean-square displacement of \qty{9}{\textmu m}, corresponding to an effective temperature of $\qty{6\times 10^{10}}{\mathrm{K}}$.
To mitigate the effects of vibrations, we implement a passive cryogenic vibration isolation system: the aluminum shield containing the trap is hung from the \qty{4}{K} stage of the dilution refrigerator via \qty{38}{\textmu m}-thick stainless steel wires and two intermediate stages. The system acts like a triple pendulum, offering isolation from horizontal ($x$ and $y$) vibrations, while the elasticity of the wires provides isolation from vertical ($z$) vibrations.
To prevent coupling of external vibrations into the experimental setup via electrical wires and the copper braids used for thermalization, we connect these wires and braids to an additional vibration isolation platform, as represented in Fig.~\ref{fig4}.
Vibrations of the experimental system could also be induced by fluctuating magnetic fields acting on the aluminum shield. To prevent this, we surround the setup by a second aluminum shield which is rigidly mounted to the cryostat. 
We typically operate at vertical ($z$) trap frequencies above \qty{200}{Hz} and initially found that the vibration isolation system attenuates vibrations at these frequencies by approximately 5 orders of magnitude. We have further improved the system by optimizing the mass and wire length for each stage, leading to an attenuation of almost 7 orders of magnitude at \qty{200}{Hz}. Theoretically, vibrations along the horizontal axis as well as librational fluctuations around the axes are suppressed even more, but in practice, we expect vibrations from the vertical axis to couple into all degrees of freedom \cite{Aldcroft1992}. More details on the vibration isolation platform are provided in \cite{Supp}.

\paragraph{Prospects for ground state cooling.---}
Now, we discuss the potential of this system to reach the ground state, which would be a first step toward accessing the quantum regime. SQUID noise can be separated into flux noise $S_{\phi\phi}$ and noise in the circulating current $S_{JJ}$, which may be partially correlated \cite{Tesche1979} and fulfill $\sqrt{S_{\phi\phi}S_{JJ}-S_{\phi J}^{2}}\geq\hbar$ \cite{Koch1981,Danilov1983}. The flux noise results in the measurement noise $\frac{S_{\phi\phi}}{\eta^2}$, while the current noise corresponds to a back action $\eta^2 S_{JJ}$, where $\eta$ is the coupling strength. Applying direct feedback with optimum gain and assuming vanishing correlations between flux and current noise, one can estimate the final phonon occupation \cite{Supp} as
\[
\bar{n} = \frac{1}{\hbar\eta^{2}}\left(k_{B}T_{0}m\gamma \tilde{L_{S}}\right) + \frac{1}{2}\left(\frac{\sqrt{S_{\phi\phi}S_{JJ}}}{\hbar}-1\right),
\]
where $\tilde{L_{S}}=\sqrt{\frac{S_{\phi\phi}}{S_{JJ}}}$ depends on the working point of the SQUID and is typically on the order of the SQUID inductance $L_{S}$. The first term describes heating from coupling to an effective thermal reservoir with temperature $T_{0}$, while the second term is the backaction limit under continuous feedback. 
We can reach a coupling strength of $\eta\approx\qty{5.5\times 10^{7}}{\mathrm{\Phi_{0}/m}}$ \cite{Supp}, so assuming $T_{0}\approx\qty{15}{\mathrm{mK}}$ and a typical SQUID inductance of $L_{S}\approx\qty{15}{\mathrm{pH}}$, we need $\gamma\approx\qty{1\times 10^{-6}}{\mathrm{s^{-1}}}$ to keep the former negligible. Ground state cooling then requires $\sqrt{S_{\phi\phi}S_{JJ}}<3\hbar$.
In addition to feedback cooling with a SQUID, we will investigate other potential avenues for ground state cooling such as coupling the levitated sphere to a flux qubit or a microwave cavity \cite{Isart2012,Schmidt2020, Shevchuk2017,Zoepfl2020}.

\paragraph{Discussion and outlook.---}
We built a platform based on the diamagnetic levitation of a superconductor in a magnetostatic trap in a dilution refrigerator at \qty{15}{mK} and demonstrate quality factors exceeding $1\times 10^{7}$ for the three c.m. modes of a \qty{5.6}{\textmu g} oscillator. The high quality factors result in excellent sensing capabilities, with achievable force (acceleration) sensitivities that are, otherwise, only reached with much smaller (larger) masses \cite{Hempston2017, Goodkind1999}. We also demonstrated simultaneous feedback cooling of all c.m. modes using either direct or parametric feedback. While we have focused, here, on a \qty{100}{\textmu m}-sized sphere, the setup allows for particles from submicron size to millimeter size to be levitated \cite{Hofer2019}.
We have identified improvements to mitigate technical issues: 
First, we plan to use a type-I superconducting sphere to avoid trapped flux and dissipation caused by flux creep, which we expect is limiting our quality factor.
Second, we can improve the magnetomechanical coupling by 4 orders of magnitude by better positioning of the pickup coil and by using a pickup coil with multiple windings matched to the input coil of the SQUID. Together with an optimized SQUID shielded from environmental noise (such that our readout is limited by the intrinsic SQUID noise), this would result in an improvement of the measurement noise floor by 6 orders of magnitude \cite{Supp}.
Third, we plan to implement a persistent current switch to improve the trap current stability. 
With these improvements, our system offers a promising approach for bringing microgram objects to the quantum regime, which opens a potential avenue for quantum-limited acceleration sensing or even probing quantum effects of gravity \cite{Kumar2020, Marletto2017, Bose2017, Aspelmeyer2022}.
\\
The data of this study are available at the Zenodo repository \cite{thedata}.

\begin{acknowledgments}
We thank G. Kirchmair, O. Romero-Isart, and A. Sanchez for discussions. We thank Matthias Rudolph for assisting with the initial setup of a SQUID measurement system. We thank F. Wulschner for setting up a helium flow cryostat and assisting with measurements during the initial phase of the project.
This work was supported by the European Union‘s Horizon 2020 research and innovation programme under Grant Agreement No. 863132 (iQLev), the European Union‘s Horizon Europe 2021-2027 Framework Programme under Grant Agreement No. 101080143 (SuperMeQ), the European Research Council (ERC CoG QLev4G and ERC Synergy QXtreme), and the Austrian Science Fund (FWF) under Project No. F40 (SFB FOQUS). We gratefully acknowledge financial support from the Deutsche Forschungsgemeinschaft (DFG, German Research Foundation) under Germany’s Excellence Strategy Grant No. EXC-2111 – 390814868. G. H. acknowledges support from the Swedish Research Council (Grant No. 2020-00381). P. S. is supported by the Alexander von Humboldt Foundation through a Feodor Lynen Fellowship. 
\end{acknowledgments}

\bibliography{ms}

\begin{thebibliography}{53}%
\makeatletter
\providecommand \@ifxundefined [1]{%
 \@ifx{#1\undefined}
}%
\providecommand \@ifnum [1]{%
 \ifnum #1\expandafter \@firstoftwo
 \else \expandafter \@secondoftwo
 \fi
}%
\providecommand \@ifx [1]{%
 \ifx #1\expandafter \@firstoftwo
 \else \expandafter \@secondoftwo
 \fi
}%
\providecommand \natexlab [1]{#1}%
\providecommand \enquote  [1]{``#1''}%
\providecommand \bibnamefont  [1]{#1}%
\providecommand \bibfnamefont [1]{#1}%
\providecommand \citenamefont [1]{#1}%
\providecommand \href@noop [0]{\@secondoftwo}%
\providecommand \href [0]{\begingroup \@sanitize@url \@href}%
\providecommand \@href[1]{\@@startlink{#1}\@@href}%
\providecommand \@@href[1]{\endgroup#1\@@endlink}%
\providecommand \@sanitize@url [0]{\catcode `\\12\catcode `\$12\catcode
  `\&12\catcode `\#12\catcode `\^12\catcode `\_12\catcode `\%12\relax}%
\providecommand \@@startlink[1]{}%
\providecommand \@@endlink[0]{}%
\providecommand \url  [0]{\begingroup\@sanitize@url \@url }%
\providecommand \@url [1]{\endgroup\@href {#1}{\urlprefix }}%
\providecommand \urlprefix  [0]{URL }%
\providecommand \Eprint [0]{\href }%
\providecommand \doibase [0]{https://doi.org/}%
\providecommand \selectlanguage [0]{\@gobble}%
\providecommand \bibinfo  [0]{\@secondoftwo}%
\providecommand \bibfield  [0]{\@secondoftwo}%
\providecommand \translation [1]{[#1]}%
\providecommand \BibitemOpen [0]{}%
\providecommand \bibitemStop [0]{}%
\providecommand \bibitemNoStop [0]{.\EOS\space}%
\providecommand \EOS [0]{\spacefactor3000\relax}%
\providecommand \BibitemShut  [1]{\csname bibitem#1\endcsname}%
\let\auto@bib@innerbib\@empty
\bibitem [{\citenamefont {Brandt}(1989)}]{Brandt1989}%
  \BibitemOpen
  \bibfield  {author} {\bibinfo {author} {\bibfnamefont {E.~H.}\ \bibnamefont
  {Brandt}},\ }\bibfield  {title} {\bibinfo {title} {Levitation in physics},\
  }\href {https://doi.org/10.1126/science.243.4889.349} {\bibfield  {journal}
  {\bibinfo  {journal} {Science}\ }\textbf {\bibinfo {volume} {243}},\ \bibinfo
  {pages} {349} (\bibinfo {year} {1989})}\BibitemShut {NoStop}%
\bibitem [{\citenamefont {Goodkind}(1999)}]{Goodkind1999}%
  \BibitemOpen
  \bibfield  {author} {\bibinfo {author} {\bibfnamefont {J.~M.}\ \bibnamefont
  {Goodkind}},\ }\bibfield  {title} {\bibinfo {title} {The superconducting
  gravimeter},\ }\href {https://doi.org/10.1063/1.1150092} {\bibfield
  {journal} {\bibinfo  {journal} {Review of Scientific Instruments}\ }\textbf
  {\bibinfo {volume} {70}},\ \bibinfo {pages} {4131} (\bibinfo {year}
  {1999})}\BibitemShut {NoStop}%
\bibitem [{\citenamefont {Romero-Isart}\ \emph {et~al.}(2012)\citenamefont
  {Romero-Isart}, \citenamefont {Clemente}, \citenamefont {Navau},
  \citenamefont {Sanchez},\ and\ \citenamefont {Cirac}}]{Isart2012}%
  \BibitemOpen
  \bibfield  {author} {\bibinfo {author} {\bibfnamefont {O.}~\bibnamefont
  {Romero-Isart}}, \bibinfo {author} {\bibfnamefont {L.}~\bibnamefont
  {Clemente}}, \bibinfo {author} {\bibfnamefont {C.}~\bibnamefont {Navau}},
  \bibinfo {author} {\bibfnamefont {A.}~\bibnamefont {Sanchez}},\ and\ \bibinfo
  {author} {\bibfnamefont {J.~I.}\ \bibnamefont {Cirac}},\ }\bibfield  {title}
  {\bibinfo {title} {Quantum magnetomechanics with levitating superconducting
  microspheres},\ }\href@noop {} {\bibfield  {journal} {\bibinfo  {journal}
  {Phys. Rev. Lett.}\ }\textbf {\bibinfo {volume} {109}},\ \bibinfo {pages}
  {147205} (\bibinfo {year} {2012})}\BibitemShut {NoStop}%
\bibitem [{\citenamefont {Cirio}\ \emph {et~al.}(2012)\citenamefont {Cirio},
  \citenamefont {Brennen},\ and\ \citenamefont {Twamley}}]{Cirio2012}%
  \BibitemOpen
  \bibfield  {author} {\bibinfo {author} {\bibfnamefont {M.}~\bibnamefont
  {Cirio}}, \bibinfo {author} {\bibfnamefont {G.~K.}\ \bibnamefont {Brennen}},\
  and\ \bibinfo {author} {\bibfnamefont {J.}~\bibnamefont {Twamley}},\
  }\bibfield  {title} {\bibinfo {title} {Quantum magnetomechanics:
  {Ultrahigh-$Q$}-levitated mechanical oscillators},\ }\href
  {https://doi.org/10.1103/PhysRevLett.109.147206} {\bibfield  {journal}
  {\bibinfo  {journal} {Phys. Rev. Lett.}\ }\textbf {\bibinfo {volume} {109}},\
  \bibinfo {pages} {147206} (\bibinfo {year} {2012})}\BibitemShut {NoStop}%
\bibitem [{\citenamefont {Vinante}\ \emph {et~al.}(2020)\citenamefont
  {Vinante}, \citenamefont {Falferi}, \citenamefont {Gasbarri}, \citenamefont
  {Setter}, \citenamefont {Timberlake},\ and\ \citenamefont
  {Ulbricht}}]{Vinante2020}%
  \BibitemOpen
  \bibfield  {author} {\bibinfo {author} {\bibfnamefont {A.}~\bibnamefont
  {Vinante}}, \bibinfo {author} {\bibfnamefont {P.}~\bibnamefont {Falferi}},
  \bibinfo {author} {\bibfnamefont {G.}~\bibnamefont {Gasbarri}}, \bibinfo
  {author} {\bibfnamefont {A.}~\bibnamefont {Setter}}, \bibinfo {author}
  {\bibfnamefont {C.}~\bibnamefont {Timberlake}},\ and\ \bibinfo {author}
  {\bibfnamefont {H.}~\bibnamefont {Ulbricht}},\ }\bibfield  {title} {\bibinfo
  {title} {Ultralow mechanical damping with {Meissner}-levitated ferromagnetic
  microparticles},\ }\bibfield  {journal} {\bibinfo  {journal} {Phys. Rev.
  Appl.}\ }\textbf {\bibinfo {volume} {13}},\ \href
  {https://doi.org/10.1103/PhysRevApplied.13.064027}
  {10.1103/PhysRevApplied.13.064027} (\bibinfo {year} {2020})\BibitemShut
  {NoStop}%
\bibitem [{\citenamefont {Gieseler}\ \emph {et~al.}(2020)\citenamefont
  {Gieseler}, \citenamefont {Kabcenell}, \citenamefont {Rosenfeld},
  \citenamefont {Schaefer}, \citenamefont {Safira}, \citenamefont {Schuetz},
  \citenamefont {Gonzalez-Ballestero}, \citenamefont {Rusconi}, \citenamefont
  {Romero-Isart},\ and\ \citenamefont {Lukin}}]{Gieseler2020}%
  \BibitemOpen
  \bibfield  {author} {\bibinfo {author} {\bibfnamefont {J.}~\bibnamefont
  {Gieseler}}, \bibinfo {author} {\bibfnamefont {A.}~\bibnamefont {Kabcenell}},
  \bibinfo {author} {\bibfnamefont {E.}~\bibnamefont {Rosenfeld}}, \bibinfo
  {author} {\bibfnamefont {J.~D.}\ \bibnamefont {Schaefer}}, \bibinfo {author}
  {\bibfnamefont {A.}~\bibnamefont {Safira}}, \bibinfo {author} {\bibfnamefont
  {M.~J.~A.}\ \bibnamefont {Schuetz}}, \bibinfo {author} {\bibfnamefont
  {C.}~\bibnamefont {Gonzalez-Ballestero}}, \bibinfo {author} {\bibfnamefont
  {C.~C.}\ \bibnamefont {Rusconi}}, \bibinfo {author} {\bibfnamefont
  {O.}~\bibnamefont {Romero-Isart}},\ and\ \bibinfo {author} {\bibfnamefont
  {M.~D.}\ \bibnamefont {Lukin}},\ }\bibfield  {title} {\bibinfo {title}
  {Single-spin magnetomechanics with levitated micromagnets},\ }\href
  {https://doi.org/10.1103/PhysRevLett.124.163604} {\bibfield  {journal}
  {\bibinfo  {journal} {Phys. Rev. Lett.}\ }\textbf {\bibinfo {volume} {124}},\
  \bibinfo {pages} {163604} (\bibinfo {year} {2020})}\BibitemShut {NoStop}%
\bibitem [{\citenamefont {Leng}\ \emph {et~al.}(2021)\citenamefont {Leng},
  \citenamefont {Li}, \citenamefont {Kong}, \citenamefont {Xie}, \citenamefont
  {Zheng}, \citenamefont {Yin}, \citenamefont {Xiong}, \citenamefont {Wu},
  \citenamefont {Duan}, \citenamefont {Du}, \citenamefont {Yin}, \citenamefont
  {Huang},\ and\ \citenamefont {Du}}]{Leng2021}%
  \BibitemOpen
  \bibfield  {author} {\bibinfo {author} {\bibfnamefont {Y.}~\bibnamefont
  {Leng}}, \bibinfo {author} {\bibfnamefont {R.}~\bibnamefont {Li}}, \bibinfo
  {author} {\bibfnamefont {X.}~\bibnamefont {Kong}}, \bibinfo {author}
  {\bibfnamefont {H.}~\bibnamefont {Xie}}, \bibinfo {author} {\bibfnamefont
  {D.}~\bibnamefont {Zheng}}, \bibinfo {author} {\bibfnamefont
  {P.}~\bibnamefont {Yin}}, \bibinfo {author} {\bibfnamefont {F.}~\bibnamefont
  {Xiong}}, \bibinfo {author} {\bibfnamefont {T.}~\bibnamefont {Wu}}, \bibinfo
  {author} {\bibfnamefont {C.-K.}\ \bibnamefont {Duan}}, \bibinfo {author}
  {\bibfnamefont {Y.}~\bibnamefont {Du}}, \bibinfo {author} {\bibfnamefont
  {Z.-q.}\ \bibnamefont {Yin}}, \bibinfo {author} {\bibfnamefont
  {P.}~\bibnamefont {Huang}},\ and\ \bibinfo {author} {\bibfnamefont
  {J.}~\bibnamefont {Du}},\ }\bibfield  {title} {\bibinfo {title} {Mechanical
  dissipation below $1\mu\mathrm{Hz}$ with a cryogenic diamagnetic levitated
  micro-oscillator},\ }\href {https://doi.org/10.1103/PhysRevApplied.15.024061}
  {\bibfield  {journal} {\bibinfo  {journal} {Phys. Rev. Applied}\ }\textbf
  {\bibinfo {volume} {15}},\ \bibinfo {pages} {024061} (\bibinfo {year}
  {2021})}\BibitemShut {NoStop}%
\bibitem [{\citenamefont {Lewandowski}\ \emph {et~al.}(2021)\citenamefont
  {Lewandowski}, \citenamefont {Knowles}, \citenamefont {Etienne},\ and\
  \citenamefont {D'Urso}}]{Urso2021}%
  \BibitemOpen
  \bibfield  {author} {\bibinfo {author} {\bibfnamefont {C.~W.}\ \bibnamefont
  {Lewandowski}}, \bibinfo {author} {\bibfnamefont {T.~D.}\ \bibnamefont
  {Knowles}}, \bibinfo {author} {\bibfnamefont {Z.~B.}\ \bibnamefont
  {Etienne}},\ and\ \bibinfo {author} {\bibfnamefont {B.}~\bibnamefont
  {D'Urso}},\ }\bibfield  {title} {\bibinfo {title} {High-sensitivity
  accelerometry with a feedback-cooled magnetically levitated microsphere},\
  }\href {https://doi.org/10.1103/PhysRevApplied.15.014050} {\bibfield
  {journal} {\bibinfo  {journal} {Phys. Rev. Applied}\ }\textbf {\bibinfo
  {volume} {15}},\ \bibinfo {pages} {014050} (\bibinfo {year}
  {2021})}\BibitemShut {NoStop}%
\bibitem [{\citenamefont {Wang}\ \emph {et~al.}(2019)\citenamefont {Wang},
  \citenamefont {Lourette}, \citenamefont {O'Kelley}, \citenamefont {Kayci},
  \citenamefont {Band}, \citenamefont {Kimball}, \citenamefont {Sushkov},\ and\
  \citenamefont {Budker}}]{Wang2019}%
  \BibitemOpen
  \bibfield  {author} {\bibinfo {author} {\bibfnamefont {T.}~\bibnamefont
  {Wang}}, \bibinfo {author} {\bibfnamefont {S.}~\bibnamefont {Lourette}},
  \bibinfo {author} {\bibfnamefont {S.~R.}\ \bibnamefont {O'Kelley}}, \bibinfo
  {author} {\bibfnamefont {M.}~\bibnamefont {Kayci}}, \bibinfo {author}
  {\bibfnamefont {Y.}~\bibnamefont {Band}}, \bibinfo {author} {\bibfnamefont
  {D.~F.~J.}\ \bibnamefont {Kimball}}, \bibinfo {author} {\bibfnamefont
  {A.~O.}\ \bibnamefont {Sushkov}},\ and\ \bibinfo {author} {\bibfnamefont
  {D.}~\bibnamefont {Budker}},\ }\bibfield  {title} {\bibinfo {title} {Dynamics
  of a ferromagnetic particle levitated over a superconductor},\ }\href
  {https://doi.org/10.1103/PhysRevApplied.11.044041} {\bibfield  {journal}
  {\bibinfo  {journal} {Phys. Rev. Applied}\ }\textbf {\bibinfo {volume}
  {11}},\ \bibinfo {pages} {044041} (\bibinfo {year} {2019})}\BibitemShut
  {NoStop}%
\bibitem [{\citenamefont {Brandt}(1988)}]{Brandt1988}%
  \BibitemOpen
  \bibfield  {author} {\bibinfo {author} {\bibfnamefont {E.~H.}\ \bibnamefont
  {Brandt}},\ }\bibfield  {title} {\bibinfo {title} {Friction in levitated
  superconductors},\ }\href@noop {} {\bibfield  {journal} {\bibinfo  {journal}
  {Applied Physics Letters}\ }\textbf {\bibinfo {volume} {53}},\ \bibinfo
  {pages} {1554} (\bibinfo {year} {1988})}\BibitemShut {NoStop}%
\bibitem [{\citenamefont {Van~Dyck}\ \emph {et~al.}(1999)\citenamefont
  {Van~Dyck}, \citenamefont {Farnham}, \citenamefont {Zafonte},\ and\
  \citenamefont {Schwinberg}}]{Van1999}%
  \BibitemOpen
  \bibfield  {author} {\bibinfo {author} {\bibfnamefont {R.~S.}\ \bibnamefont
  {Van~Dyck}}, \bibinfo {author} {\bibfnamefont {D.~L.}\ \bibnamefont
  {Farnham}}, \bibinfo {author} {\bibfnamefont {S.~L.}\ \bibnamefont
  {Zafonte}},\ and\ \bibinfo {author} {\bibfnamefont {P.~B.}\ \bibnamefont
  {Schwinberg}},\ }\bibfield  {title} {\bibinfo {title} {Ultrastable
  superconducting magnet system for a penning trap mass spectrometer},\ }\href
  {https://doi.org/10.1063/1.1149649} {\bibfield  {journal} {\bibinfo
  {journal} {Review of Scientific Instruments}\ }\textbf {\bibinfo {volume}
  {70}},\ \bibinfo {pages} {1665} (\bibinfo {year} {1999})}\BibitemShut
  {NoStop}%
\bibitem [{\citenamefont {Britton}\ \emph {et~al.}(2016)\citenamefont
  {Britton}, \citenamefont {Bohnet}, \citenamefont {Sawyer}, \citenamefont
  {Uys}, \citenamefont {Biercuk},\ and\ \citenamefont
  {Bollinger}}]{Britton2016}%
  \BibitemOpen
  \bibfield  {author} {\bibinfo {author} {\bibfnamefont {J.~W.}\ \bibnamefont
  {Britton}}, \bibinfo {author} {\bibfnamefont {J.~G.}\ \bibnamefont {Bohnet}},
  \bibinfo {author} {\bibfnamefont {B.~C.}\ \bibnamefont {Sawyer}}, \bibinfo
  {author} {\bibfnamefont {H.}~\bibnamefont {Uys}}, \bibinfo {author}
  {\bibfnamefont {M.~J.}\ \bibnamefont {Biercuk}},\ and\ \bibinfo {author}
  {\bibfnamefont {J.~J.}\ \bibnamefont {Bollinger}},\ }\bibfield  {title}
  {\bibinfo {title} {Vibration-induced field fluctuations in a superconducting
  magnet},\ }\href {https://doi.org/10.1103/PhysRevA.93.062511} {\bibfield
  {journal} {\bibinfo  {journal} {Phys. Rev. A}\ }\textbf {\bibinfo {volume}
  {93}},\ \bibinfo {pages} {062511} (\bibinfo {year} {2016})}\BibitemShut
  {NoStop}%
\bibitem [{\citenamefont {Latorre}\ \emph {et~al.}(2020)\citenamefont
  {Latorre}, \citenamefont {Hofer}, \citenamefont {Rudolph},\ and\
  \citenamefont {Wieczorek}}]{Latorre2020}%
  \BibitemOpen
  \bibfield  {author} {\bibinfo {author} {\bibfnamefont {M.~G.}\ \bibnamefont
  {Latorre}}, \bibinfo {author} {\bibfnamefont {J.}~\bibnamefont {Hofer}},
  \bibinfo {author} {\bibfnamefont {M.}~\bibnamefont {Rudolph}},\ and\ \bibinfo
  {author} {\bibfnamefont {W.}~\bibnamefont {Wieczorek}},\ }\bibfield  {title}
  {\bibinfo {title} {Chip-based superconducting traps for levitation of
  micrometer-sized particles in the {Meissner} state},\ }\href
  {https://doi.org/10.1088/1361-6668/aba6e1} {\bibfield  {journal} {\bibinfo
  {journal} {Superconductor Science and Technology}\ }\textbf {\bibinfo
  {volume} {33}},\ \bibinfo {pages} {105002} (\bibinfo {year}
  {2020})}\BibitemShut {NoStop}%
\bibitem [{\citenamefont {Latorre}\ \emph {et~al.}(2022)\citenamefont
  {Latorre}, \citenamefont {Paradkar}, \citenamefont {Hambraeus}, \citenamefont
  {Higgins},\ and\ \citenamefont {Wieczorek}}]{Latorre2022}%
  \BibitemOpen
  \bibfield  {author} {\bibinfo {author} {\bibfnamefont {M.~G.}\ \bibnamefont
  {Latorre}}, \bibinfo {author} {\bibfnamefont {A.}~\bibnamefont {Paradkar}},
  \bibinfo {author} {\bibfnamefont {D.}~\bibnamefont {Hambraeus}}, \bibinfo
  {author} {\bibfnamefont {G.}~\bibnamefont {Higgins}},\ and\ \bibinfo {author}
  {\bibfnamefont {W.}~\bibnamefont {Wieczorek}},\ }\bibfield  {title} {\bibinfo
  {title} {A chip-based superconducting magnetic trap for levitating
  superconducting microparticles},\ }\href
  {https://doi.org/10.1109/TASC.2022.3147730} {\bibfield  {journal} {\bibinfo
  {journal} {IEEE Transactions on Applied Superconductivity}\ }\textbf
  {\bibinfo {volume} {32}},\ \bibinfo {pages} {1} (\bibinfo {year}
  {2022})}\BibitemShut {NoStop}%
\bibitem [{\citenamefont {Gutierrez~Latorre}\ \emph {et~al.}(2023)\citenamefont
  {Gutierrez~Latorre}, \citenamefont {Higgins}, \citenamefont {Paradkar},
  \citenamefont {Bauch},\ and\ \citenamefont {Wieczorek}}]{Latorre2022_2}%
  \BibitemOpen
  \bibfield  {author} {\bibinfo {author} {\bibfnamefont {M.}~\bibnamefont
  {Gutierrez~Latorre}}, \bibinfo {author} {\bibfnamefont {G.}~\bibnamefont
  {Higgins}}, \bibinfo {author} {\bibfnamefont {A.}~\bibnamefont {Paradkar}},
  \bibinfo {author} {\bibfnamefont {T.}~\bibnamefont {Bauch}},\ and\ \bibinfo
  {author} {\bibfnamefont {W.}~\bibnamefont {Wieczorek}},\ }\bibfield  {title}
  {\bibinfo {title} {Superconducting microsphere magnetically levitated in an
  anharmonic potential with integrated magnetic readout},\ }\href
  {https://doi.org/10.1103/PhysRevApplied.19.054047} {\bibfield  {journal}
  {\bibinfo  {journal} {Phys. Rev. Appl.}\ }\textbf {\bibinfo {volume} {19}},\
  \bibinfo {pages} {054047} (\bibinfo {year} {2023})}\BibitemShut {NoStop}%
\bibitem [{\citenamefont {Pino}\ \emph {et~al.}(2018)\citenamefont {Pino},
  \citenamefont {Prat-Camps}, \citenamefont {Sinha}, \citenamefont
  {Venkatesh},\ and\ \citenamefont {Romero-Isart}}]{Pino2018}%
  \BibitemOpen
  \bibfield  {author} {\bibinfo {author} {\bibfnamefont {H.}~\bibnamefont
  {Pino}}, \bibinfo {author} {\bibfnamefont {J.}~\bibnamefont {Prat-Camps}},
  \bibinfo {author} {\bibfnamefont {K.}~\bibnamefont {Sinha}}, \bibinfo
  {author} {\bibfnamefont {B.~P.}\ \bibnamefont {Venkatesh}},\ and\ \bibinfo
  {author} {\bibfnamefont {O.}~\bibnamefont {Romero-Isart}},\ }\bibfield
  {title} {\bibinfo {title} {On-chip quantum interference of a superconducting
  microsphere},\ }\href {https://doi.org/10.1088/2058-9565/aa9d15} {\bibfield
  {journal} {\bibinfo  {journal} {Quantum Science and Technology}\ }\textbf
  {\bibinfo {volume} {3}},\ \bibinfo {pages} {025001} (\bibinfo {year}
  {2018})}\BibitemShut {NoStop}%
\bibitem [{\citenamefont {Delic}\ \emph {et~al.}(2020)\citenamefont {Delic},
  \citenamefont {Reisenbauer}, \citenamefont {Dare}, \citenamefont {Grass},
  \citenamefont {{Vuletic Vladan}}, \citenamefont {Kiesel},\ and\ \citenamefont
  {Aspelmeyer}}]{Delic2020}%
  \BibitemOpen
  \bibfield  {author} {\bibinfo {author} {\bibfnamefont {U.}~\bibnamefont
  {Delic}}, \bibinfo {author} {\bibfnamefont {M.}~\bibnamefont {Reisenbauer}},
  \bibinfo {author} {\bibfnamefont {K.}~\bibnamefont {Dare}}, \bibinfo {author}
  {\bibfnamefont {D.}~\bibnamefont {Grass}}, \bibinfo {author} {\bibnamefont
  {{Vuletic Vladan}}}, \bibinfo {author} {\bibfnamefont {N.}~\bibnamefont
  {Kiesel}},\ and\ \bibinfo {author} {\bibfnamefont {M.}~\bibnamefont
  {Aspelmeyer}},\ }\bibfield  {title} {\bibinfo {title} {Cooling of a levitated
  nanoparticle to the motional quantum ground state},\ }\href@noop {}
  {\bibfield  {journal} {\bibinfo  {journal} {Science}\ }\textbf {\bibinfo
  {volume} {367}},\ \bibinfo {pages} {892} (\bibinfo {year}
  {2020})}\BibitemShut {NoStop}%
\bibitem [{\citenamefont {Magrini}\ \emph {et~al.}(2021)\citenamefont
  {Magrini}, \citenamefont {Rosenzweig}, \citenamefont {Bach}, \citenamefont
  {Deutschmann-Olek}, \citenamefont {Hofer}, \citenamefont {Hong},
  \citenamefont {Kiesel}, \citenamefont {Kugi},\ and\ \citenamefont
  {Aspelmeyer}}]{Magrini2021}%
  \BibitemOpen
  \bibfield  {author} {\bibinfo {author} {\bibfnamefont {L.}~\bibnamefont
  {Magrini}}, \bibinfo {author} {\bibfnamefont {P.}~\bibnamefont {Rosenzweig}},
  \bibinfo {author} {\bibfnamefont {C.}~\bibnamefont {Bach}}, \bibinfo {author}
  {\bibfnamefont {A.}~\bibnamefont {Deutschmann-Olek}}, \bibinfo {author}
  {\bibfnamefont {S.~G.}\ \bibnamefont {Hofer}}, \bibinfo {author}
  {\bibfnamefont {S.}~\bibnamefont {Hong}}, \bibinfo {author} {\bibfnamefont
  {N.}~\bibnamefont {Kiesel}}, \bibinfo {author} {\bibfnamefont
  {A.}~\bibnamefont {Kugi}},\ and\ \bibinfo {author} {\bibfnamefont
  {M.}~\bibnamefont {Aspelmeyer}},\ }\bibfield  {title} {\bibinfo {title}
  {Real-time optimal quantum control of mechanical motion at room
  temperature},\ }\href {https://doi.org/10.1038/s41586-021-03602-3} {\bibfield
   {journal} {\bibinfo  {journal} {Nature}\ }\textbf {\bibinfo {volume}
  {595}},\ \bibinfo {pages} {373} (\bibinfo {year} {2021})}\BibitemShut
  {NoStop}%
\bibitem [{\citenamefont {Tebbenjohanns}\ \emph {et~al.}(2021)\citenamefont
  {Tebbenjohanns}, \citenamefont {Mattana}, \citenamefont {Rossi},
  \citenamefont {Frimmer},\ and\ \citenamefont {Novotny}}]{Tebbenjohanns2021}%
  \BibitemOpen
  \bibfield  {author} {\bibinfo {author} {\bibfnamefont {F.}~\bibnamefont
  {Tebbenjohanns}}, \bibinfo {author} {\bibfnamefont {M.~L.}\ \bibnamefont
  {Mattana}}, \bibinfo {author} {\bibfnamefont {M.}~\bibnamefont {Rossi}},
  \bibinfo {author} {\bibfnamefont {M.}~\bibnamefont {Frimmer}},\ and\ \bibinfo
  {author} {\bibfnamefont {L.}~\bibnamefont {Novotny}},\ }\bibfield  {title}
  {\bibinfo {title} {Quantum control of a nanoparticle optically levitated in
  cryogenic free space},\ }\href {https://doi.org/10.1038/s41586-021-03617-w}
  {\bibfield  {journal} {\bibinfo  {journal} {Nature}\ }\textbf {\bibinfo
  {volume} {595}},\ \bibinfo {pages} {378} (\bibinfo {year}
  {2021})}\BibitemShut {NoStop}%
\bibitem [{\citenamefont {Hofer}\ and\ \citenamefont
  {Aspelmeyer}(2019)}]{Hofer2019}%
  \BibitemOpen
  \bibfield  {author} {\bibinfo {author} {\bibfnamefont {J.}~\bibnamefont
  {Hofer}}\ and\ \bibinfo {author} {\bibfnamefont {M.}~\bibnamefont
  {Aspelmeyer}},\ }\bibfield  {title} {\bibinfo {title} {{Analytic solutions to
  the Maxwell{\textendash}London equations and levitation force for a
  superconducting sphere in a quadrupole field}},\ }\href
  {https://doi.org/10.1088/1402-4896/ab0c44} {\bibfield  {journal} {\bibinfo
  {journal} {Physica Scripta}\ }\textbf {\bibinfo {volume} {94}},\ \bibinfo
  {pages} {125508} (\bibinfo {year} {2019})}\BibitemShut {NoStop}%
\bibitem [{\citenamefont {Navau}\ \emph {et~al.}(2021)\citenamefont {Navau},
  \citenamefont {Minniberger}, \citenamefont {Trupke},\ and\ \citenamefont
  {Sanchez}}]{Navau2021}%
  \BibitemOpen
  \bibfield  {author} {\bibinfo {author} {\bibfnamefont {C.}~\bibnamefont
  {Navau}}, \bibinfo {author} {\bibfnamefont {S.}~\bibnamefont {Minniberger}},
  \bibinfo {author} {\bibfnamefont {M.}~\bibnamefont {Trupke}},\ and\ \bibinfo
  {author} {\bibfnamefont {A.}~\bibnamefont {Sanchez}},\ }\bibfield  {title}
  {\bibinfo {title} {Levitation of superconducting microrings for quantum
  magnetomechanics},\ }\href {https://doi.org/10.1103/PhysRevB.103.174436}
  {\bibfield  {journal} {\bibinfo  {journal} {Phys. Rev. B}\ }\textbf {\bibinfo
  {volume} {103}},\ \bibinfo {pages} {174436} (\bibinfo {year}
  {2021})}\BibitemShut {NoStop}%
\bibitem [{\citenamefont {Monteiro}\ \emph {et~al.}(2017)\citenamefont
  {Monteiro}, \citenamefont {Ghosh}, \citenamefont {Fine},\ and\ \citenamefont
  {Moore}}]{Moore2017}%
  \BibitemOpen
  \bibfield  {author} {\bibinfo {author} {\bibfnamefont {F.}~\bibnamefont
  {Monteiro}}, \bibinfo {author} {\bibfnamefont {S.}~\bibnamefont {Ghosh}},
  \bibinfo {author} {\bibfnamefont {A.~G.}\ \bibnamefont {Fine}},\ and\
  \bibinfo {author} {\bibfnamefont {D.~C.}\ \bibnamefont {Moore}},\ }\bibfield
  {title} {\bibinfo {title} {Optical levitation of 10-ng spheres with nano-$g$
  acceleration sensitivity},\ }\href
  {https://doi.org/10.1103/PhysRevA.96.063841} {\bibfield  {journal} {\bibinfo
  {journal} {Phys. Rev. A}\ }\textbf {\bibinfo {volume} {96}},\ \bibinfo
  {pages} {063841} (\bibinfo {year} {2017})}\BibitemShut {NoStop}%
\bibitem [{\citenamefont {Francis C.~Moon}(1995)}]{Moon1995}%
  \BibitemOpen
  \bibfield  {author} {\bibinfo {author} {\bibfnamefont {P.-Z.~C.}\
  \bibnamefont {Francis C.~Moon}},\ }\href@noop {} {\emph {\bibinfo {title}
  {Superconducting Levitation: Applications to Bearings and Magnetic
  Transportation}}}\ (\bibinfo  {publisher} {Wiley-VCH},\ \bibinfo {year}
  {1995})\BibitemShut {NoStop}%
\bibitem [{\citenamefont {van Waarde}(2016)}]{Waarde2016}%
  \BibitemOpen
  \bibfield  {author} {\bibinfo {author} {\bibfnamefont {B.}~\bibnamefont {van
  Waarde}},\ }\emph {\bibinfo {title} {The lead zeppelin : a force sensor
  without a handle}},\ \href@noop {} {\bibinfo {type} {{Ph.D.} thesis}},\
  \bibinfo  {school} {Leiden University} (\bibinfo {year} {2016})\BibitemShut
  {NoStop}%
\bibitem [{Sup()}]{Supp}%
  \BibitemOpen
  \href@noop {} {\bibinfo {title} {{See supplemental material, which includes
  Refs. \cite{Neuhaus2017, Hinkle1990, Martinis1985, Crocker1996,
  Awschalom1988, Carelli1998, Wheeler1928, Penny2021, Martinis1983,
  Falferi2008, Gehm1998}, for more information on the derivation of trap
  parameters, the effects of trap current fluctuations, the vibration isolation
  system, estimations of different contributions to the damping, the
  calibration of the SQUID signal as well as more details on ground state
  cooling.}}}\BibitemShut {Stop}%
\bibitem [{\citenamefont {Livingston}(1963)}]{Livingston1963}%
  \BibitemOpen
  \bibfield  {author} {\bibinfo {author} {\bibfnamefont {J.~D.}\ \bibnamefont
  {Livingston}},\ }\bibfield  {title} {\bibinfo {title} {Magnetic properties of
  superconducting lead-base alloys},\ }\href
  {https://doi.org/10.1103/PhysRev.129.1943} {\bibfield  {journal} {\bibinfo
  {journal} {Phys. Rev.}\ }\textbf {\bibinfo {volume} {129}},\ \bibinfo {pages}
  {1943} (\bibinfo {year} {1963})}\BibitemShut {NoStop}%
\bibitem [{\citenamefont {Storm}\ \emph {et~al.}(2020)\citenamefont {Storm},
  \citenamefont {Kieler},\ and\ \citenamefont {K{\"o}rber}}]{Storm2020}%
  \BibitemOpen
  \bibfield  {author} {\bibinfo {author} {\bibfnamefont {J.-H.}\ \bibnamefont
  {Storm}}, \bibinfo {author} {\bibfnamefont {O.}~\bibnamefont {Kieler}},\ and\
  \bibinfo {author} {\bibfnamefont {R.}~\bibnamefont {K{\"o}rber}},\ }\bibfield
   {title} {\bibinfo {title} {{Towards ultrasensitive SQUIDs based on
  submicrometer-sized Josephson junctions}},\ }\href
  {https://doi.org/10.1109/TASC.2020.2989630} {\bibfield  {journal} {\bibinfo
  {journal} {IEEE Transactions on Applied Superconductivity}\ }\textbf
  {\bibinfo {volume} {30}},\ \bibinfo {pages} {1} (\bibinfo {year}
  {2020})}\BibitemShut {NoStop}%
\bibitem [{\citenamefont {Nemoshkalenko}\ \emph {et~al.}(1990)\citenamefont
  {Nemoshkalenko}, \citenamefont {Brandt}, \citenamefont {Kordyuk},\ and\
  \citenamefont {Nikitin}}]{Nemoshkalenko1990}%
  \BibitemOpen
  \bibfield  {author} {\bibinfo {author} {\bibfnamefont {V.~V.}\ \bibnamefont
  {Nemoshkalenko}}, \bibinfo {author} {\bibfnamefont {E.~H.}\ \bibnamefont
  {Brandt}}, \bibinfo {author} {\bibfnamefont {A.~A.}\ \bibnamefont
  {Kordyuk}},\ and\ \bibinfo {author} {\bibfnamefont {B.~G.}\ \bibnamefont
  {Nikitin}},\ }\bibfield  {title} {\bibinfo {title} {{Dynamics of a permanent
  magnet levitating above a high-$\mathrm{T_c}$ superconductor}},\ }\href@noop
  {} {\bibfield  {journal} {\bibinfo  {journal} {Physica C}\ }\textbf {\bibinfo
  {volume} {170}},\ \bibinfo {pages} {481–485} (\bibinfo {year}
  {1990})}\BibitemShut {NoStop}%
\bibitem [{\citenamefont {Kubo}(1966)}]{Kubo1966}%
  \BibitemOpen
  \bibfield  {author} {\bibinfo {author} {\bibfnamefont {R.}~\bibnamefont
  {Kubo}},\ }\bibfield  {title} {\bibinfo {title} {The fluctuation-dissipation
  theorem},\ }\href {https://doi.org/10.1088/0034-4885/29/1/306} {\bibfield
  {journal} {\bibinfo  {journal} {Reports on Progress in Physics}\ }\textbf
  {\bibinfo {volume} {29}},\ \bibinfo {pages} {255} (\bibinfo {year}
  {1966})}\BibitemShut {NoStop}%
\bibitem [{\citenamefont {Aldcroft}\ \emph {et~al.}(1992)\citenamefont
  {Aldcroft}, \citenamefont {Michelson}, \citenamefont {Taber},\ and\
  \citenamefont {McLoughlin}}]{Aldcroft1992}%
  \BibitemOpen
  \bibfield  {author} {\bibinfo {author} {\bibfnamefont {T.~L.}\ \bibnamefont
  {Aldcroft}}, \bibinfo {author} {\bibfnamefont {P.~F.}\ \bibnamefont
  {Michelson}}, \bibinfo {author} {\bibfnamefont {R.~C.}\ \bibnamefont
  {Taber}},\ and\ \bibinfo {author} {\bibfnamefont {F.~A.}\ \bibnamefont
  {McLoughlin}},\ }\bibfield  {title} {\bibinfo {title}
  {Six‐degree‐of‐freedom vibration isolation systems with application to
  resonant‐mass gravitational radiation detectors},\ }\href
  {https://doi.org/10.1063/1.1143277} {\bibfield  {journal} {\bibinfo
  {journal} {Review of Scientific Instruments}\ }\textbf {\bibinfo {volume}
  {63}},\ \bibinfo {pages} {3815} (\bibinfo {year} {1992})}\BibitemShut
  {NoStop}%
\bibitem [{\citenamefont {Tesche}\ and\ \citenamefont
  {Clarke}(1979)}]{Tesche1979}%
  \BibitemOpen
  \bibfield  {author} {\bibinfo {author} {\bibfnamefont {C.~D.}\ \bibnamefont
  {Tesche}}\ and\ \bibinfo {author} {\bibfnamefont {J.}~\bibnamefont
  {Clarke}},\ }\bibfield  {title} {\bibinfo {title} {{dc SQUID: Current
  noise}},\ }\href {https://doi.org/10.1007/BF00119197} {\bibfield  {journal}
  {\bibinfo  {journal} {Journal of Low Temperature Physics}\ }\textbf {\bibinfo
  {volume} {37}},\ \bibinfo {pages} {397} (\bibinfo {year} {1979})}\BibitemShut
  {NoStop}%
\bibitem [{\citenamefont {Koch}\ \emph {et~al.}(1981)\citenamefont {Koch},
  \citenamefont {Van~Harlingen},\ and\ \citenamefont {Clarke}}]{Koch1981}%
  \BibitemOpen
  \bibfield  {author} {\bibinfo {author} {\bibfnamefont {R.~H.}\ \bibnamefont
  {Koch}}, \bibinfo {author} {\bibfnamefont {D.~J.}\ \bibnamefont
  {Van~Harlingen}},\ and\ \bibinfo {author} {\bibfnamefont {J.}~\bibnamefont
  {Clarke}},\ }\bibfield  {title} {\bibinfo {title} {Quantum noise theory for
  the dc {SQUID}},\ }\href {https://doi.org/10.1063/1.92345} {\bibfield
  {journal} {\bibinfo  {journal} {Applied Physics Letters}\ }\textbf {\bibinfo
  {volume} {38}},\ \bibinfo {pages} {380} (\bibinfo {year} {1981})}\BibitemShut
  {NoStop}%
\bibitem [{\citenamefont {Danilov}\ \emph {et~al.}(1983)\citenamefont
  {Danilov}, \citenamefont {Likharev},\ and\ \citenamefont
  {Zorin}}]{Danilov1983}%
  \BibitemOpen
  \bibfield  {author} {\bibinfo {author} {\bibfnamefont {V.}~\bibnamefont
  {Danilov}}, \bibinfo {author} {\bibfnamefont {K.}~\bibnamefont {Likharev}},\
  and\ \bibinfo {author} {\bibfnamefont {A.}~\bibnamefont {Zorin}},\ }\bibfield
   {title} {\bibinfo {title} {Quantum noise in squids},\ }\href
  {https://doi.org/10.1109/TMAG.1983.1062489} {\bibfield  {journal} {\bibinfo
  {journal} {IEEE Transactions on Magnetics}\ }\textbf {\bibinfo {volume}
  {19}},\ \bibinfo {pages} {572} (\bibinfo {year} {1983})}\BibitemShut
  {NoStop}%
\bibitem [{\citenamefont {Schmidt}\ \emph {et~al.}(2020)\citenamefont
  {Schmidt}, \citenamefont {T.~Amawi}, \citenamefont {Pogorzalek},
  \citenamefont {Deppe}, \citenamefont {Marx}, \citenamefont {Gross},\ and\
  \citenamefont {Huebl}}]{Schmidt2020}%
  \BibitemOpen
  \bibfield  {author} {\bibinfo {author} {\bibfnamefont {P.}~\bibnamefont
  {Schmidt}}, \bibinfo {author} {\bibfnamefont {M.}~\bibnamefont {T.~Amawi}},
  \bibinfo {author} {\bibfnamefont {S.}~\bibnamefont {Pogorzalek}}, \bibinfo
  {author} {\bibfnamefont {F.}~\bibnamefont {Deppe}}, \bibinfo {author}
  {\bibfnamefont {A.}~\bibnamefont {Marx}}, \bibinfo {author} {\bibfnamefont
  {R.}~\bibnamefont {Gross}},\ and\ \bibinfo {author} {\bibfnamefont
  {H.}~\bibnamefont {Huebl}},\ }\bibfield  {title} {\bibinfo {title}
  {Sideband-resolved resonator electromechanics based on a nonlinear
  {Josephson} inductance probed on the single-photon level},\ }\href
  {https://doi.org/10.1038/s42005-020-00501-3} {\bibfield  {journal} {\bibinfo
  {journal} {Communications Physics}\ }\textbf {\bibinfo {volume} {3}},\
  \bibinfo {pages} {233} (\bibinfo {year} {2020})}\BibitemShut {NoStop}%
\bibitem [{\citenamefont {Shevchuk}\ \emph {et~al.}(2017)\citenamefont
  {Shevchuk}, \citenamefont {Steele},\ and\ \citenamefont
  {Blanter}}]{Shevchuk2017}%
  \BibitemOpen
  \bibfield  {author} {\bibinfo {author} {\bibfnamefont {O.}~\bibnamefont
  {Shevchuk}}, \bibinfo {author} {\bibfnamefont {G.~A.}\ \bibnamefont
  {Steele}},\ and\ \bibinfo {author} {\bibfnamefont {Y.~M.}\ \bibnamefont
  {Blanter}},\ }\bibfield  {title} {\bibinfo {title} {Strong and tunable
  couplings in flux-mediated optomechanics},\ }\href
  {https://doi.org/10.1103/PhysRevB.96.014508} {\bibfield  {journal} {\bibinfo
  {journal} {Phys. Rev. B}\ }\textbf {\bibinfo {volume} {96}},\ \bibinfo
  {pages} {014508} (\bibinfo {year} {2017})}\BibitemShut {NoStop}%
\bibitem [{\citenamefont {Zoepfl}\ \emph {et~al.}(2020)\citenamefont {Zoepfl},
  \citenamefont {Juan}, \citenamefont {Schneider},\ and\ \citenamefont
  {Kirchmair}}]{Zoepfl2020}%
  \BibitemOpen
  \bibfield  {author} {\bibinfo {author} {\bibfnamefont {D.}~\bibnamefont
  {Zoepfl}}, \bibinfo {author} {\bibfnamefont {M.~L.}\ \bibnamefont {Juan}},
  \bibinfo {author} {\bibfnamefont {C.~M.~F.}\ \bibnamefont {Schneider}},\ and\
  \bibinfo {author} {\bibfnamefont {G.}~\bibnamefont {Kirchmair}},\ }\bibfield
  {title} {\bibinfo {title} {Single-photon cooling in microwave
  magnetomechanics},\ }\href {https://doi.org/10.1103/PhysRevLett.125.023601}
  {\bibfield  {journal} {\bibinfo  {journal} {Phys. Rev. Lett.}\ }\textbf
  {\bibinfo {volume} {125}},\ \bibinfo {pages} {023601} (\bibinfo {year}
  {2020})}\BibitemShut {NoStop}%
\bibitem [{\citenamefont {Hempston}\ \emph {et~al.}(2017)\citenamefont
  {Hempston}, \citenamefont {Vovrosh}, \citenamefont {Toroš}, \citenamefont
  {Winstone}, \citenamefont {Rashid},\ and\ \citenamefont
  {Ulbricht}}]{Hempston2017}%
  \BibitemOpen
  \bibfield  {author} {\bibinfo {author} {\bibfnamefont {D.}~\bibnamefont
  {Hempston}}, \bibinfo {author} {\bibfnamefont {J.}~\bibnamefont {Vovrosh}},
  \bibinfo {author} {\bibfnamefont {M.}~\bibnamefont {Toroš}}, \bibinfo
  {author} {\bibfnamefont {G.}~\bibnamefont {Winstone}}, \bibinfo {author}
  {\bibfnamefont {M.}~\bibnamefont {Rashid}},\ and\ \bibinfo {author}
  {\bibfnamefont {H.}~\bibnamefont {Ulbricht}},\ }\bibfield  {title} {\bibinfo
  {title} {Force sensing with an optically levitated charged nanoparticle},\
  }\href {https://doi.org/10.1063/1.4993555} {\bibfield  {journal} {\bibinfo
  {journal} {Applied Physics Letters}\ }\textbf {\bibinfo {volume} {111}},\
  \bibinfo {pages} {133111} (\bibinfo {year} {2017})}\BibitemShut {NoStop}%
\bibitem [{\citenamefont {Kumar}\ and\ \citenamefont
  {Plenio}(2020)}]{Kumar2020}%
  \BibitemOpen
  \bibfield  {author} {\bibinfo {author} {\bibfnamefont {S.~P.}\ \bibnamefont
  {Kumar}}\ and\ \bibinfo {author} {\bibfnamefont {M.~B.}\ \bibnamefont
  {Plenio}},\ }\bibfield  {title} {\bibinfo {title} {On quantum gravity tests
  with composite particles},\ }\href
  {https://doi.org/10.1038/s41467-020-17518-5} {\bibfield  {journal} {\bibinfo
  {journal} {Nature Communications}\ }\textbf {\bibinfo {volume} {11}},\
  \bibinfo {pages} {3900} (\bibinfo {year} {2020})}\BibitemShut {NoStop}%
\bibitem [{\citenamefont {Marletto}\ and\ \citenamefont
  {Vedral}(2017)}]{Marletto2017}%
  \BibitemOpen
  \bibfield  {author} {\bibinfo {author} {\bibfnamefont {C.}~\bibnamefont
  {Marletto}}\ and\ \bibinfo {author} {\bibfnamefont {V.}~\bibnamefont
  {Vedral}},\ }\bibfield  {title} {\bibinfo {title} {Gravitationally induced
  entanglement between two massive particles is sufficient evidence of quantum
  effects in gravity},\ }\href {https://doi.org/10.1103/PhysRevLett.119.240402}
  {\bibfield  {journal} {\bibinfo  {journal} {Phys. Rev. Lett.}\ }\textbf
  {\bibinfo {volume} {119}},\ \bibinfo {pages} {240402} (\bibinfo {year}
  {2017})}\BibitemShut {NoStop}%
\bibitem [{\citenamefont {Bose}\ \emph {et~al.}(2017)\citenamefont {Bose},
  \citenamefont {Mazumdar}, \citenamefont {Morley}, \citenamefont {Ulbricht},
  \citenamefont {Toro\ifmmode~\check{s}\else \v{s}\fi{}}, \citenamefont
  {Paternostro}, \citenamefont {Geraci}, \citenamefont {Barker}, \citenamefont
  {Kim},\ and\ \citenamefont {Milburn}}]{Bose2017}%
  \BibitemOpen
  \bibfield  {author} {\bibinfo {author} {\bibfnamefont {S.}~\bibnamefont
  {Bose}}, \bibinfo {author} {\bibfnamefont {A.}~\bibnamefont {Mazumdar}},
  \bibinfo {author} {\bibfnamefont {G.~W.}\ \bibnamefont {Morley}}, \bibinfo
  {author} {\bibfnamefont {H.}~\bibnamefont {Ulbricht}}, \bibinfo {author}
  {\bibfnamefont {M.}~\bibnamefont {Toro\ifmmode~\check{s}\else \v{s}\fi{}}},
  \bibinfo {author} {\bibfnamefont {M.}~\bibnamefont {Paternostro}}, \bibinfo
  {author} {\bibfnamefont {A.~A.}\ \bibnamefont {Geraci}}, \bibinfo {author}
  {\bibfnamefont {P.~F.}\ \bibnamefont {Barker}}, \bibinfo {author}
  {\bibfnamefont {M.~S.}\ \bibnamefont {Kim}},\ and\ \bibinfo {author}
  {\bibfnamefont {G.}~\bibnamefont {Milburn}},\ }\bibfield  {title} {\bibinfo
  {title} {Spin entanglement witness for quantum gravity},\ }\href
  {https://doi.org/10.1103/PhysRevLett.119.240401} {\bibfield  {journal}
  {\bibinfo  {journal} {Phys. Rev. Lett.}\ }\textbf {\bibinfo {volume} {119}},\
  \bibinfo {pages} {240401} (\bibinfo {year} {2017})}\BibitemShut {NoStop}%
\bibitem [{\citenamefont {Aspelmeyer}(2022)}]{Aspelmeyer2022}%
  \BibitemOpen
  \bibfield  {author} {\bibinfo {author} {\bibfnamefont {M.}~\bibnamefont
  {Aspelmeyer}},\ }\href {https://doi.org/10.48550/ARXIV.2203.05587} {\bibinfo
  {title} {How to avoid the appearance of a classical world in gravity
  experiments}} (\bibinfo {year} {2022}),\ \Eprint
  {https://arxiv.org/abs/2203.05587} {arXiv:2203.05587 [quant-ph]} \BibitemShut
  {NoStop}%
\bibitem [{\citenamefont {{Hofer et al.}}(2023)}]{thedata}%
  \BibitemOpen
  \bibfield  {author} {\bibinfo {author} {\bibfnamefont {J.}~\bibnamefont
  {{Hofer et al.}}},\ }\bibfield  {title} {\bibinfo {title} {Data used in the
  article "{High-Q Magnetic Levitation and Control of Superconducting
  Microspheres at Millikelvin Temperatures}"},\ }\bibfield  {journal} {\bibinfo
   {journal} {Zenodo,}\ }\href {https://doi.org/10.5281/zenodo.7837944}
  {10.5281/zenodo.7837944} (\bibinfo {year} {2023})\BibitemShut {NoStop}%
\bibitem [{\citenamefont {Neuhaus}\ \emph {et~al.}(2017)\citenamefont
  {Neuhaus}, \citenamefont {Metzdorff}, \citenamefont {Chua}, \citenamefont
  {Jacqmin}, \citenamefont {Briant}, \citenamefont {Heidmann}, \citenamefont
  {Cohadon},\ and\ \citenamefont {Deléglise}}]{Neuhaus2017}%
  \BibitemOpen
  \bibfield  {author} {\bibinfo {author} {\bibfnamefont {L.}~\bibnamefont
  {Neuhaus}}, \bibinfo {author} {\bibfnamefont {R.}~\bibnamefont {Metzdorff}},
  \bibinfo {author} {\bibfnamefont {S.}~\bibnamefont {Chua}}, \bibinfo {author}
  {\bibfnamefont {T.}~\bibnamefont {Jacqmin}}, \bibinfo {author} {\bibfnamefont
  {T.}~\bibnamefont {Briant}}, \bibinfo {author} {\bibfnamefont
  {A.}~\bibnamefont {Heidmann}}, \bibinfo {author} {\bibfnamefont {P.-F.}\
  \bibnamefont {Cohadon}},\ and\ \bibinfo {author} {\bibfnamefont
  {S.}~\bibnamefont {Deléglise}},\ }\bibfield  {title} {\bibinfo {title}
  {{PyRPL (Python Red Pitaya Lockbox) — An open-source software package for
  FPGA-controlled quantum optics experiments}},\ }in\ \href
  {https://doi.org/10.1109/CLEOE-EQEC.2017.8087380} {\emph {\bibinfo
  {booktitle} {2017 Conference on Lasers and Electro-Optics Europe European
  Quantum Electronics Conference (CLEO/Europe-EQEC)}}}\ (\bibinfo {year}
  {2017})\ pp.\ \bibinfo {pages} {1--1}\BibitemShut {NoStop}%
\bibitem [{\citenamefont {Hinkle}\ and\ \citenamefont
  {Kendall}(1990)}]{Hinkle1990}%
  \BibitemOpen
  \bibfield  {author} {\bibinfo {author} {\bibfnamefont {L.~D.}\ \bibnamefont
  {Hinkle}}\ and\ \bibinfo {author} {\bibfnamefont {B.~R.~F.}\ \bibnamefont
  {Kendall}},\ }\bibfield  {title} {\bibinfo {title} {Pressure‐dependent
  damping of a particle levitated in vacuum},\ }\href
  {https://doi.org/10.1116/1.576672} {\bibfield  {journal} {\bibinfo  {journal}
  {Journal of Vacuum Science \& Technology A}\ }\textbf {\bibinfo {volume}
  {8}},\ \bibinfo {pages} {2802} (\bibinfo {year} {1990})}\BibitemShut
  {NoStop}%
\bibitem [{\citenamefont {Martinis}\ and\ \citenamefont
  {Clarke}(1985)}]{Martinis1985}%
  \BibitemOpen
  \bibfield  {author} {\bibinfo {author} {\bibfnamefont {J.~M.}\ \bibnamefont
  {Martinis}}\ and\ \bibinfo {author} {\bibfnamefont {J.}~\bibnamefont
  {Clarke}},\ }\bibfield  {title} {\bibinfo {title} {Signal and noise theory
  for a dc {SQUID} amplifier},\ }\href {https://doi.org/10.1007/BF00681633}
  {\bibfield  {journal} {\bibinfo  {journal} {Journal of Low Temperature
  Physics}\ }\textbf {\bibinfo {volume} {61}},\ \bibinfo {pages} {227}
  (\bibinfo {year} {1985})}\BibitemShut {NoStop}%
\bibitem [{\citenamefont {Crocker}\ and\ \citenamefont
  {Grier}(1996)}]{Crocker1996}%
  \BibitemOpen
  \bibfield  {author} {\bibinfo {author} {\bibfnamefont {J.~C.}\ \bibnamefont
  {Crocker}}\ and\ \bibinfo {author} {\bibfnamefont {D.~G.}\ \bibnamefont
  {Grier}},\ }\bibfield  {title} {\bibinfo {title} {Methods of digital video
  microscopy for colloidal studies},\ }\href
  {https://doi.org/https://doi.org/10.1006/jcis.1996.0217} {\bibfield
  {journal} {\bibinfo  {journal} {Journal of Colloid and Interface Science}\
  }\textbf {\bibinfo {volume} {179}},\ \bibinfo {pages} {298} (\bibinfo {year}
  {1996})}\BibitemShut {NoStop}%
\bibitem [{\citenamefont {Awschalom}\ \emph {et~al.}(1988)\citenamefont
  {Awschalom}, \citenamefont {Rozen}, \citenamefont {Ketchen}, \citenamefont
  {Gallagher}, \citenamefont {Kleinsasser}, \citenamefont {Sandstrom},\ and\
  \citenamefont {Bumble}}]{Awschalom1988}%
  \BibitemOpen
  \bibfield  {author} {\bibinfo {author} {\bibfnamefont {D.~D.}\ \bibnamefont
  {Awschalom}}, \bibinfo {author} {\bibfnamefont {J.~R.}\ \bibnamefont
  {Rozen}}, \bibinfo {author} {\bibfnamefont {M.~B.}\ \bibnamefont {Ketchen}},
  \bibinfo {author} {\bibfnamefont {W.~J.}\ \bibnamefont {Gallagher}}, \bibinfo
  {author} {\bibfnamefont {A.~W.}\ \bibnamefont {Kleinsasser}}, \bibinfo
  {author} {\bibfnamefont {R.~L.}\ \bibnamefont {Sandstrom}},\ and\ \bibinfo
  {author} {\bibfnamefont {B.}~\bibnamefont {Bumble}},\ }\bibfield  {title}
  {\bibinfo {title} {Low‐noise modular microsusceptometer using nearly
  quantum limited dc {SQUIDs}},\ }\href {https://doi.org/10.1063/1.100291}
  {\bibfield  {journal} {\bibinfo  {journal} {Applied Physics Letters}\
  }\textbf {\bibinfo {volume} {53}},\ \bibinfo {pages} {2108} (\bibinfo {year}
  {1988})}\BibitemShut {NoStop}%
\bibitem [{\citenamefont {Carelli}\ \emph {et~al.}(1998)\citenamefont
  {Carelli}, \citenamefont {Castellano}, \citenamefont {Torrioli},\ and\
  \citenamefont {Leoni}}]{Carelli1998}%
  \BibitemOpen
  \bibfield  {author} {\bibinfo {author} {\bibfnamefont {P.}~\bibnamefont
  {Carelli}}, \bibinfo {author} {\bibfnamefont {M.~G.}\ \bibnamefont
  {Castellano}}, \bibinfo {author} {\bibfnamefont {G.}~\bibnamefont
  {Torrioli}},\ and\ \bibinfo {author} {\bibfnamefont {R.}~\bibnamefont
  {Leoni}},\ }\bibfield  {title} {\bibinfo {title} {Low noise multiwasher
  superconducting interferometer},\ }\href {https://doi.org/10.1063/1.121444}
  {\bibfield  {journal} {\bibinfo  {journal} {Applied Physics Letters}\
  }\textbf {\bibinfo {volume} {72}},\ \bibinfo {pages} {115} (\bibinfo {year}
  {1998})}\BibitemShut {NoStop}%
\bibitem [{\citenamefont {Wheeler}(1928)}]{Wheeler1928}%
  \BibitemOpen
  \bibfield  {author} {\bibinfo {author} {\bibfnamefont {H.}~\bibnamefont
  {Wheeler}},\ }\bibfield  {title} {\bibinfo {title} {Simple inductance
  formulas for radio coils},\ }\href
  {https://doi.org/10.1109/JRPROC.1928.221309} {\bibfield  {journal} {\bibinfo
  {journal} {Proceedings of the Institute of Radio Engineers}\ }\textbf
  {\bibinfo {volume} {16}},\ \bibinfo {pages} {1398} (\bibinfo {year}
  {1928})}\BibitemShut {NoStop}%
\bibitem [{\citenamefont {Penny}\ \emph {et~al.}(2021)\citenamefont {Penny},
  \citenamefont {Pontin},\ and\ \citenamefont {Barker}}]{Penny2021}%
  \BibitemOpen
  \bibfield  {author} {\bibinfo {author} {\bibfnamefont {T.~W.}\ \bibnamefont
  {Penny}}, \bibinfo {author} {\bibfnamefont {A.}~\bibnamefont {Pontin}},\ and\
  \bibinfo {author} {\bibfnamefont {P.~F.}\ \bibnamefont {Barker}},\ }\bibfield
   {title} {\bibinfo {title} {{Performance and limits of feedback cooling
  methods for levitated oscillators: A direct comparison}},\ }\href
  {https://doi.org/10.1103/PhysRevA.104.023502} {\bibfield  {journal} {\bibinfo
   {journal} {Phys. Rev. A}\ }\textbf {\bibinfo {volume} {104}},\ \bibinfo
  {pages} {023502} (\bibinfo {year} {2021})}\BibitemShut {NoStop}%
\bibitem [{\citenamefont {Martinis}\ and\ \citenamefont
  {Clarke}(1983)}]{Martinis1983}%
  \BibitemOpen
  \bibfield  {author} {\bibinfo {author} {\bibfnamefont {J.}~\bibnamefont
  {Martinis}}\ and\ \bibinfo {author} {\bibfnamefont {J.}~\bibnamefont
  {Clarke}},\ }\bibfield  {title} {\bibinfo {title} {Measurements of current
  noise in dc {SQUIDS}},\ }\href {https://doi.org/10.1109/TMAG.1983.1062411}
  {\bibfield  {journal} {\bibinfo  {journal} {IEEE Transactions on Magnetics}\
  }\textbf {\bibinfo {volume} {19}},\ \bibinfo {pages} {446} (\bibinfo {year}
  {1983})}\BibitemShut {NoStop}%
\bibitem [{\citenamefont {Falferi}\ \emph {et~al.}(2008)\citenamefont
  {Falferi}, \citenamefont {Bonaldi}, \citenamefont {Cerdonio}, \citenamefont
  {Mezzena}, \citenamefont {Prodi}, \citenamefont {Vinante},\ and\
  \citenamefont {Vitale}}]{Falferi2008}%
  \BibitemOpen
  \bibfield  {author} {\bibinfo {author} {\bibfnamefont {P.}~\bibnamefont
  {Falferi}}, \bibinfo {author} {\bibfnamefont {M.}~\bibnamefont {Bonaldi}},
  \bibinfo {author} {\bibfnamefont {M.}~\bibnamefont {Cerdonio}}, \bibinfo
  {author} {\bibfnamefont {R.}~\bibnamefont {Mezzena}}, \bibinfo {author}
  {\bibfnamefont {G.~A.}\ \bibnamefont {Prodi}}, \bibinfo {author}
  {\bibfnamefont {A.}~\bibnamefont {Vinante}},\ and\ \bibinfo {author}
  {\bibfnamefont {S.}~\bibnamefont {Vitale}},\ }\bibfield  {title} {\bibinfo
  {title} {10$\hbar$ superconducting quantum interference device amplifier for
  acoustic gravitational wave detectors},\ }\href
  {https://doi.org/10.1063/1.3002321} {\bibfield  {journal} {\bibinfo
  {journal} {Applied Physics Letters}\ }\textbf {\bibinfo {volume} {93}},\
  \bibinfo {pages} {172506} (\bibinfo {year} {2008})}\BibitemShut {NoStop}%
\bibitem [{\citenamefont {Gehm}\ \emph {et~al.}(1998)\citenamefont {Gehm},
  \citenamefont {O'Hara}, \citenamefont {Savard},\ and\ \citenamefont
  {Thomas}}]{Gehm1998}%
  \BibitemOpen
  \bibfield  {author} {\bibinfo {author} {\bibfnamefont {M.~E.}\ \bibnamefont
  {Gehm}}, \bibinfo {author} {\bibfnamefont {K.~M.}\ \bibnamefont {O'Hara}},
  \bibinfo {author} {\bibfnamefont {T.~A.}\ \bibnamefont {Savard}},\ and\
  \bibinfo {author} {\bibfnamefont {J.~E.}\ \bibnamefont {Thomas}},\ }\bibfield
   {title} {\bibinfo {title} {Dynamics of noise-induced heating in atom
  traps},\ }\href {https://doi.org/10.1103/PhysRevA.58.3914} {\bibfield
  {journal} {\bibinfo  {journal} {Phys. Rev. A}\ }\textbf {\bibinfo {volume}
  {58}},\ \bibinfo {pages} {3914} (\bibinfo {year} {1998})}\BibitemShut
  {NoStop}%
\end{thebibliography}%


\begin{thebibliography}{19}%
\makeatletter
\providecommand \@ifxundefined [1]{%
 \@ifx{#1\undefined}
}%
\providecommand \@ifnum [1]{%
 \ifnum #1\expandafter \@firstoftwo
 \else \expandafter \@secondoftwo
 \fi
}%
\providecommand \@ifx [1]{%
 \ifx #1\expandafter \@firstoftwo
 \else \expandafter \@secondoftwo
 \fi
}%
\providecommand \natexlab [1]{#1}%
\providecommand \enquote  [1]{``#1''}%
\providecommand \bibnamefont  [1]{#1}%
\providecommand \bibfnamefont [1]{#1}%
\providecommand \citenamefont [1]{#1}%
\providecommand \href@noop [0]{\@secondoftwo}%
\providecommand \href [0]{\begingroup \@sanitize@url \@href}%
\providecommand \@href[1]{\@@startlink{#1}\@@href}%
\providecommand \@@href[1]{\endgroup#1\@@endlink}%
\providecommand \@sanitize@url [0]{\catcode `\\12\catcode `\$12\catcode
  `\&12\catcode `\#12\catcode `\^12\catcode `\_12\catcode `\%12\relax}%
\providecommand \@@startlink[1]{}%
\providecommand \@@endlink[0]{}%
\providecommand \url  [0]{\begingroup\@sanitize@url \@url }%
\providecommand \@url [1]{\endgroup\@href {#1}{\urlprefix }}%
\providecommand \urlprefix  [0]{URL }%
\providecommand \Eprint [0]{\href }%
\providecommand \doibase [0]{https://doi.org/}%
\providecommand \selectlanguage [0]{\@gobble}%
\providecommand \bibinfo  [0]{\@secondoftwo}%
\providecommand \bibfield  [0]{\@secondoftwo}%
\providecommand \translation [1]{[#1]}%
\providecommand \BibitemOpen [0]{}%
\providecommand \bibitemStop [0]{}%
\providecommand \bibitemNoStop [0]{.\EOS\space}%
\providecommand \EOS [0]{\spacefactor3000\relax}%
\providecommand \BibitemShut  [1]{\csname bibitem#1\endcsname}%
\let\auto@bib@innerbib\@empty
\bibitem [{\citenamefont {Hofer}\ and\ \citenamefont
  {Aspelmeyer}(2019)}]{Hofer2019}%
  \BibitemOpen
  \bibfield  {author} {\bibinfo {author} {\bibfnamefont {J.}~\bibnamefont
  {Hofer}}\ and\ \bibinfo {author} {\bibfnamefont {M.}~\bibnamefont
  {Aspelmeyer}},\ }\bibfield  {title} {\bibinfo {title} {{Analytic solutions to
  the Maxwell{\textendash}London equations and levitation force for a
  superconducting sphere in a quadrupole field}},\ }\href
  {https://doi.org/10.1088/1402-4896/ab0c44} {\bibfield  {journal} {\bibinfo
  {journal} {Physica Scripta}\ }\textbf {\bibinfo {volume} {94}},\ \bibinfo
  {pages} {125508} (\bibinfo {year} {2019})}\BibitemShut {NoStop}%
\bibitem [{\citenamefont {Livingston}(1963)}]{Livingston1963}%
  \BibitemOpen
  \bibfield  {author} {\bibinfo {author} {\bibfnamefont {J.~D.}\ \bibnamefont
  {Livingston}},\ }\bibfield  {title} {\bibinfo {title} {Magnetic properties of
  superconducting lead-base alloys},\ }\href
  {https://doi.org/10.1103/PhysRev.129.1943} {\bibfield  {journal} {\bibinfo
  {journal} {Phys. Rev.}\ }\textbf {\bibinfo {volume} {129}},\ \bibinfo {pages}
  {1943} (\bibinfo {year} {1963})}\BibitemShut {NoStop}%
\bibitem [{\citenamefont {Van~Dyck}\ \emph {et~al.}(1999)\citenamefont
  {Van~Dyck}, \citenamefont {Farnham}, \citenamefont {Zafonte},\ and\
  \citenamefont {Schwinberg}}]{Van1999}%
  \BibitemOpen
  \bibfield  {author} {\bibinfo {author} {\bibfnamefont {R.~S.}\ \bibnamefont
  {Van~Dyck}}, \bibinfo {author} {\bibfnamefont {D.~L.}\ \bibnamefont
  {Farnham}}, \bibinfo {author} {\bibfnamefont {S.~L.}\ \bibnamefont
  {Zafonte}},\ and\ \bibinfo {author} {\bibfnamefont {P.~B.}\ \bibnamefont
  {Schwinberg}},\ }\bibfield  {title} {\bibinfo {title} {Ultrastable
  superconducting magnet system for a penning trap mass spectrometer},\ }\href
  {https://doi.org/10.1063/1.1149649} {\bibfield  {journal} {\bibinfo
  {journal} {Review of Scientific Instruments}\ }\textbf {\bibinfo {volume}
  {70}},\ \bibinfo {pages} {1665} (\bibinfo {year} {1999})}\BibitemShut
  {NoStop}%
\bibitem [{\citenamefont {Neuhaus}\ \emph {et~al.}(2017)\citenamefont
  {Neuhaus}, \citenamefont {Metzdorff}, \citenamefont {Chua}, \citenamefont
  {Jacqmin}, \citenamefont {Briant}, \citenamefont {Heidmann}, \citenamefont
  {Cohadon},\ and\ \citenamefont {Deléglise}}]{Neuhaus2017}%
  \BibitemOpen
  \bibfield  {author} {\bibinfo {author} {\bibfnamefont {L.}~\bibnamefont
  {Neuhaus}}, \bibinfo {author} {\bibfnamefont {R.}~\bibnamefont {Metzdorff}},
  \bibinfo {author} {\bibfnamefont {S.}~\bibnamefont {Chua}}, \bibinfo {author}
  {\bibfnamefont {T.}~\bibnamefont {Jacqmin}}, \bibinfo {author} {\bibfnamefont
  {T.}~\bibnamefont {Briant}}, \bibinfo {author} {\bibfnamefont
  {A.}~\bibnamefont {Heidmann}}, \bibinfo {author} {\bibfnamefont {P.-F.}\
  \bibnamefont {Cohadon}},\ and\ \bibinfo {author} {\bibfnamefont
  {S.}~\bibnamefont {Deléglise}},\ }\bibfield  {title} {\bibinfo {title}
  {{PyRPL (Python Red Pitaya Lockbox) — An open-source software package for
  FPGA-controlled quantum optics experiments}},\ }in\ \href
  {https://doi.org/10.1109/CLEOE-EQEC.2017.8087380} {\emph {\bibinfo
  {booktitle} {2017 Conference on Lasers and Electro-Optics Europe European
  Quantum Electronics Conference (CLEO/Europe-EQEC)}}}\ (\bibinfo {year}
  {2017})\ pp.\ \bibinfo {pages} {1--1}\BibitemShut {NoStop}%
\bibitem [{\citenamefont {Aldcroft}\ \emph {et~al.}(1992)\citenamefont
  {Aldcroft}, \citenamefont {Michelson}, \citenamefont {Taber},\ and\
  \citenamefont {McLoughlin}}]{Aldcroft1992}%
  \BibitemOpen
  \bibfield  {author} {\bibinfo {author} {\bibfnamefont {T.~L.}\ \bibnamefont
  {Aldcroft}}, \bibinfo {author} {\bibfnamefont {P.~F.}\ \bibnamefont
  {Michelson}}, \bibinfo {author} {\bibfnamefont {R.~C.}\ \bibnamefont
  {Taber}},\ and\ \bibinfo {author} {\bibfnamefont {F.~A.}\ \bibnamefont
  {McLoughlin}},\ }\bibfield  {title} {\bibinfo {title}
  {Six‐degree‐of‐freedom vibration isolation systems with application to
  resonant‐mass gravitational radiation detectors},\ }\href
  {https://doi.org/10.1063/1.1143277} {\bibfield  {journal} {\bibinfo
  {journal} {Review of Scientific Instruments}\ }\textbf {\bibinfo {volume}
  {63}},\ \bibinfo {pages} {3815} (\bibinfo {year} {1992})}\BibitemShut
  {NoStop}%
\bibitem [{\citenamefont {Hinkle}\ and\ \citenamefont
  {Kendall}(1990)}]{Hinkle1990}%
  \BibitemOpen
  \bibfield  {author} {\bibinfo {author} {\bibfnamefont {L.~D.}\ \bibnamefont
  {Hinkle}}\ and\ \bibinfo {author} {\bibfnamefont {B.~R.~F.}\ \bibnamefont
  {Kendall}},\ }\bibfield  {title} {\bibinfo {title} {Pressure‐dependent
  damping of a particle levitated in vacuum},\ }\href
  {https://doi.org/10.1116/1.576672} {\bibfield  {journal} {\bibinfo  {journal}
  {Journal of Vacuum Science \& Technology A}\ }\textbf {\bibinfo {volume}
  {8}},\ \bibinfo {pages} {2802} (\bibinfo {year} {1990})}\BibitemShut
  {NoStop}%
\bibitem [{\citenamefont {Martinis}\ and\ \citenamefont
  {Clarke}(1985)}]{Martinis1985}%
  \BibitemOpen
  \bibfield  {author} {\bibinfo {author} {\bibfnamefont {J.~M.}\ \bibnamefont
  {Martinis}}\ and\ \bibinfo {author} {\bibfnamefont {J.}~\bibnamefont
  {Clarke}},\ }\bibfield  {title} {\bibinfo {title} {Signal and noise theory
  for a dc {SQUID} amplifier},\ }\href {https://doi.org/10.1007/BF00681633}
  {\bibfield  {journal} {\bibinfo  {journal} {Journal of Low Temperature
  Physics}\ }\textbf {\bibinfo {volume} {61}},\ \bibinfo {pages} {227}
  (\bibinfo {year} {1985})}\BibitemShut {NoStop}%
\bibitem [{\citenamefont {Crocker}\ and\ \citenamefont
  {Grier}(1996)}]{Crocker1996}%
  \BibitemOpen
  \bibfield  {author} {\bibinfo {author} {\bibfnamefont {J.~C.}\ \bibnamefont
  {Crocker}}\ and\ \bibinfo {author} {\bibfnamefont {D.~G.}\ \bibnamefont
  {Grier}},\ }\bibfield  {title} {\bibinfo {title} {Methods of digital video
  microscopy for colloidal studies},\ }\href
  {https://doi.org/https://doi.org/10.1006/jcis.1996.0217} {\bibfield
  {journal} {\bibinfo  {journal} {Journal of Colloid and Interface Science}\
  }\textbf {\bibinfo {volume} {179}},\ \bibinfo {pages} {298} (\bibinfo {year}
  {1996})}\BibitemShut {NoStop}%
\bibitem [{\citenamefont {Awschalom}\ \emph {et~al.}(1988)\citenamefont
  {Awschalom}, \citenamefont {Rozen}, \citenamefont {Ketchen}, \citenamefont
  {Gallagher}, \citenamefont {Kleinsasser}, \citenamefont {Sandstrom},\ and\
  \citenamefont {Bumble}}]{Awschalom1988}%
  \BibitemOpen
  \bibfield  {author} {\bibinfo {author} {\bibfnamefont {D.~D.}\ \bibnamefont
  {Awschalom}}, \bibinfo {author} {\bibfnamefont {J.~R.}\ \bibnamefont
  {Rozen}}, \bibinfo {author} {\bibfnamefont {M.~B.}\ \bibnamefont {Ketchen}},
  \bibinfo {author} {\bibfnamefont {W.~J.}\ \bibnamefont {Gallagher}}, \bibinfo
  {author} {\bibfnamefont {A.~W.}\ \bibnamefont {Kleinsasser}}, \bibinfo
  {author} {\bibfnamefont {R.~L.}\ \bibnamefont {Sandstrom}},\ and\ \bibinfo
  {author} {\bibfnamefont {B.}~\bibnamefont {Bumble}},\ }\bibfield  {title}
  {\bibinfo {title} {Low‐noise modular microsusceptometer using nearly
  quantum limited dc {SQUIDs}},\ }\href {https://doi.org/10.1063/1.100291}
  {\bibfield  {journal} {\bibinfo  {journal} {Applied Physics Letters}\
  }\textbf {\bibinfo {volume} {53}},\ \bibinfo {pages} {2108} (\bibinfo {year}
  {1988})}\BibitemShut {NoStop}%
\bibitem [{\citenamefont {Carelli}\ \emph {et~al.}(1998)\citenamefont
  {Carelli}, \citenamefont {Castellano}, \citenamefont {Torrioli},\ and\
  \citenamefont {Leoni}}]{Carelli1998}%
  \BibitemOpen
  \bibfield  {author} {\bibinfo {author} {\bibfnamefont {P.}~\bibnamefont
  {Carelli}}, \bibinfo {author} {\bibfnamefont {M.~G.}\ \bibnamefont
  {Castellano}}, \bibinfo {author} {\bibfnamefont {G.}~\bibnamefont
  {Torrioli}},\ and\ \bibinfo {author} {\bibfnamefont {R.}~\bibnamefont
  {Leoni}},\ }\bibfield  {title} {\bibinfo {title} {Low noise multiwasher
  superconducting interferometer},\ }\href {https://doi.org/10.1063/1.121444}
  {\bibfield  {journal} {\bibinfo  {journal} {Applied Physics Letters}\
  }\textbf {\bibinfo {volume} {72}},\ \bibinfo {pages} {115} (\bibinfo {year}
  {1998})}\BibitemShut {NoStop}%
\bibitem [{\citenamefont {Wheeler}(1928)}]{Wheeler1928}%
  \BibitemOpen
  \bibfield  {author} {\bibinfo {author} {\bibfnamefont {H.}~\bibnamefont
  {Wheeler}},\ }\bibfield  {title} {\bibinfo {title} {Simple inductance
  formulas for radio coils},\ }\href
  {https://doi.org/10.1109/JRPROC.1928.221309} {\bibfield  {journal} {\bibinfo
  {journal} {Proceedings of the Institute of Radio Engineers}\ }\textbf
  {\bibinfo {volume} {16}},\ \bibinfo {pages} {1398} (\bibinfo {year}
  {1928})}\BibitemShut {NoStop}%
\bibitem [{\citenamefont {Koch}\ \emph {et~al.}(1981)\citenamefont {Koch},
  \citenamefont {Van~Harlingen},\ and\ \citenamefont {Clarke}}]{Koch1981}%
  \BibitemOpen
  \bibfield  {author} {\bibinfo {author} {\bibfnamefont {R.~H.}\ \bibnamefont
  {Koch}}, \bibinfo {author} {\bibfnamefont {D.~J.}\ \bibnamefont
  {Van~Harlingen}},\ and\ \bibinfo {author} {\bibfnamefont {J.}~\bibnamefont
  {Clarke}},\ }\bibfield  {title} {\bibinfo {title} {Quantum noise theory for
  the dc {SQUID}},\ }\href {https://doi.org/10.1063/1.92345} {\bibfield
  {journal} {\bibinfo  {journal} {Applied Physics Letters}\ }\textbf {\bibinfo
  {volume} {38}},\ \bibinfo {pages} {380} (\bibinfo {year} {1981})}\BibitemShut
  {NoStop}%
\bibitem [{\citenamefont {Danilov}\ \emph {et~al.}(1983)\citenamefont
  {Danilov}, \citenamefont {Likharev},\ and\ \citenamefont
  {Zorin}}]{Danilov1983}%
  \BibitemOpen
  \bibfield  {author} {\bibinfo {author} {\bibfnamefont {V.}~\bibnamefont
  {Danilov}}, \bibinfo {author} {\bibfnamefont {K.}~\bibnamefont {Likharev}},\
  and\ \bibinfo {author} {\bibfnamefont {A.}~\bibnamefont {Zorin}},\ }\bibfield
   {title} {\bibinfo {title} {Quantum noise in squids},\ }\href
  {https://doi.org/10.1109/TMAG.1983.1062489} {\bibfield  {journal} {\bibinfo
  {journal} {IEEE Transactions on Magnetics}\ }\textbf {\bibinfo {volume}
  {19}},\ \bibinfo {pages} {572} (\bibinfo {year} {1983})}\BibitemShut
  {NoStop}%
\bibitem [{\citenamefont {Tesche}\ and\ \citenamefont
  {Clarke}(1979)}]{Tesche1979}%
  \BibitemOpen
  \bibfield  {author} {\bibinfo {author} {\bibfnamefont {C.~D.}\ \bibnamefont
  {Tesche}}\ and\ \bibinfo {author} {\bibfnamefont {J.}~\bibnamefont
  {Clarke}},\ }\bibfield  {title} {\bibinfo {title} {{dc SQUID: Current
  noise}},\ }\href {https://doi.org/10.1007/BF00119197} {\bibfield  {journal}
  {\bibinfo  {journal} {Journal of Low Temperature Physics}\ }\textbf {\bibinfo
  {volume} {37}},\ \bibinfo {pages} {397} (\bibinfo {year} {1979})}\BibitemShut
  {NoStop}%
\bibitem [{\citenamefont {Penny}\ \emph {et~al.}(2021)\citenamefont {Penny},
  \citenamefont {Pontin},\ and\ \citenamefont {Barker}}]{Penny2021}%
  \BibitemOpen
  \bibfield  {author} {\bibinfo {author} {\bibfnamefont {T.~W.}\ \bibnamefont
  {Penny}}, \bibinfo {author} {\bibfnamefont {A.}~\bibnamefont {Pontin}},\ and\
  \bibinfo {author} {\bibfnamefont {P.~F.}\ \bibnamefont {Barker}},\ }\bibfield
   {title} {\bibinfo {title} {{Performance and limits of feedback cooling
  methods for levitated oscillators: A direct comparison}},\ }\href
  {https://doi.org/10.1103/PhysRevA.104.023502} {\bibfield  {journal} {\bibinfo
   {journal} {Phys. Rev. A}\ }\textbf {\bibinfo {volume} {104}},\ \bibinfo
  {pages} {023502} (\bibinfo {year} {2021})}\BibitemShut {NoStop}%
\bibitem [{\citenamefont {Martinis}\ and\ \citenamefont
  {Clarke}(1983)}]{Martinis1983}%
  \BibitemOpen
  \bibfield  {author} {\bibinfo {author} {\bibfnamefont {J.}~\bibnamefont
  {Martinis}}\ and\ \bibinfo {author} {\bibfnamefont {J.}~\bibnamefont
  {Clarke}},\ }\bibfield  {title} {\bibinfo {title} {Measurements of current
  noise in dc {SQUIDS}},\ }\href {https://doi.org/10.1109/TMAG.1983.1062411}
  {\bibfield  {journal} {\bibinfo  {journal} {IEEE Transactions on Magnetics}\
  }\textbf {\bibinfo {volume} {19}},\ \bibinfo {pages} {446} (\bibinfo {year}
  {1983})}\BibitemShut {NoStop}%
\bibitem [{\citenamefont {Falferi}\ \emph {et~al.}(2008)\citenamefont
  {Falferi}, \citenamefont {Bonaldi}, \citenamefont {Cerdonio}, \citenamefont
  {Mezzena}, \citenamefont {Prodi}, \citenamefont {Vinante},\ and\
  \citenamefont {Vitale}}]{Falferi2008}%
  \BibitemOpen
  \bibfield  {author} {\bibinfo {author} {\bibfnamefont {P.}~\bibnamefont
  {Falferi}}, \bibinfo {author} {\bibfnamefont {M.}~\bibnamefont {Bonaldi}},
  \bibinfo {author} {\bibfnamefont {M.}~\bibnamefont {Cerdonio}}, \bibinfo
  {author} {\bibfnamefont {R.}~\bibnamefont {Mezzena}}, \bibinfo {author}
  {\bibfnamefont {G.~A.}\ \bibnamefont {Prodi}}, \bibinfo {author}
  {\bibfnamefont {A.}~\bibnamefont {Vinante}},\ and\ \bibinfo {author}
  {\bibfnamefont {S.}~\bibnamefont {Vitale}},\ }\bibfield  {title} {\bibinfo
  {title} {10$\hbar$ superconducting quantum interference device amplifier for
  acoustic gravitational wave detectors},\ }\href
  {https://doi.org/10.1063/1.3002321} {\bibfield  {journal} {\bibinfo
  {journal} {Applied Physics Letters}\ }\textbf {\bibinfo {volume} {93}},\
  \bibinfo {pages} {172506} (\bibinfo {year} {2008})}\BibitemShut {NoStop}%
\bibitem [{\citenamefont {Gehm}\ \emph {et~al.}(1998)\citenamefont {Gehm},
  \citenamefont {O'Hara}, \citenamefont {Savard},\ and\ \citenamefont
  {Thomas}}]{Gehm1998}%
  \BibitemOpen
  \bibfield  {author} {\bibinfo {author} {\bibfnamefont {M.~E.}\ \bibnamefont
  {Gehm}}, \bibinfo {author} {\bibfnamefont {K.~M.}\ \bibnamefont {O'Hara}},
  \bibinfo {author} {\bibfnamefont {T.~A.}\ \bibnamefont {Savard}},\ and\
  \bibinfo {author} {\bibfnamefont {J.~E.}\ \bibnamefont {Thomas}},\ }\bibfield
   {title} {\bibinfo {title} {Dynamics of noise-induced heating in atom
  traps},\ }\href {https://doi.org/10.1103/PhysRevA.58.3914} {\bibfield
  {journal} {\bibinfo  {journal} {Phys. Rev. A}\ }\textbf {\bibinfo {volume}
  {58}},\ \bibinfo {pages} {3914} (\bibinfo {year} {1998})}\BibitemShut
  {NoStop}%
\bibitem [{\citenamefont {Britton}\ \emph {et~al.}(2016)\citenamefont
  {Britton}, \citenamefont {Bohnet}, \citenamefont {Sawyer}, \citenamefont
  {Uys}, \citenamefont {Biercuk},\ and\ \citenamefont
  {Bollinger}}]{Britton2016}%
  \BibitemOpen
  \bibfield  {author} {\bibinfo {author} {\bibfnamefont {J.~W.}\ \bibnamefont
  {Britton}}, \bibinfo {author} {\bibfnamefont {J.~G.}\ \bibnamefont {Bohnet}},
  \bibinfo {author} {\bibfnamefont {B.~C.}\ \bibnamefont {Sawyer}}, \bibinfo
  {author} {\bibfnamefont {H.}~\bibnamefont {Uys}}, \bibinfo {author}
  {\bibfnamefont {M.~J.}\ \bibnamefont {Biercuk}},\ and\ \bibinfo {author}
  {\bibfnamefont {J.~J.}\ \bibnamefont {Bollinger}},\ }\bibfield  {title}
  {\bibinfo {title} {Vibration-induced field fluctuations in a superconducting
  magnet},\ }\href {https://doi.org/10.1103/PhysRevA.93.062511} {\bibfield
  {journal} {\bibinfo  {journal} {Phys. Rev. A}\ }\textbf {\bibinfo {volume}
  {93}},\ \bibinfo {pages} {062511} (\bibinfo {year} {2016})}\BibitemShut
  {NoStop}%
\end{thebibliography}%

\end{document}


\title{Supplemental material: High-Q magnetic levitation and control of superconducting microspheres at millikelvin temperatures}

\author{J. Hofer}
\email[]{joachim.hofer@univie.ac.at}
\affiliation{University of Vienna, Faculty of Physics, Vienna Center for Quantum Science and Technology (VCQ), A-1090 Vienna, Austria}
\affiliation{Institute for Quantum Optics and Quantum Information (IQOQI), Austrian Academy of Sciences, A-1090 Vienna, Austria}
\author{R. Gross}
\affiliation{Walther-Meißner-Institut, Bayerische Akademie der Wissenschaften, D-85748 Garching, Germany}
\affiliation{Physik-Department, Technische Universität München, D-85748 Garching, Germany}
\affiliation{Munich Center for Quantum Science and Technology (MCQST), D-80799 München, Germany}
\author{G. Higgins}
\affiliation{Institute for Quantum Optics and Quantum Information (IQOQI), Austrian Academy of Sciences, A-1090 Vienna, Austria}
\affiliation{Department of Microtechnology and Nanoscience (MC2), Chalmers University of Technology, S-412 96 Gothenburg, Sweden}
\author{H. Huebl}
\affiliation{Walther-Meißner-Institut, Bayerische Akademie der Wissenschaften, D-85748 Garching, Germany}
\affiliation{Physik-Department, Technische Universität München, D-85748 Garching, Germany}
\affiliation{Munich Center for Quantum Science and Technology (MCQST), D-80799 München, Germany}
\author{O. F. Kieler}
\affiliation{Physikalisch-Technische Bundesanstalt (PTB), D-38116 Braunschweig, Germany}
\author{R. Kleiner}
\affiliation{Physikalisches Institut, Center for Quantum Science (CQ) and LISA\textsuperscript{+}, University of Tuebingen, D-72076 Tuebingen, Germany}
\author{D. Koelle}
\affiliation{Physikalisches Institut, Center for Quantum Science (CQ) and LISA\textsuperscript{+}, University of Tuebingen, D-72076 Tuebingen, Germany}
\author{P. Schmidt}
\affiliation{Institute for Quantum Optics and Quantum Information (IQOQI), Austrian Academy of Sciences, A-1090 Vienna, Austria}
\author{J. A. Slater}
\altaffiliation{Currently at: QuTech, Delft University of Technology, Delft, The Netherlands}
\affiliation{University of Vienna, Faculty of Physics, Vienna Center for Quantum Science and Technology (VCQ), A-1090 Vienna, Austria}
\author{M. Trupke}
\affiliation{University of Vienna, Faculty of Physics, Vienna Center for Quantum Science and Technology (VCQ), A-1090 Vienna, Austria}
\author{K. Uhl}
\affiliation{Physikalisches Institut, Center for Quantum Science (CQ) and LISA\textsuperscript{+}, University of Tuebingen, D-72076 Tuebingen, Germany}
\author{T. Weimann}
\affiliation{Physikalisch-Technische Bundesanstalt (PTB), D-38116 Braunschweig, Germany}
\author{W. Wieczorek}
\affiliation{University of Vienna, Faculty of Physics, Vienna Center for Quantum Science and Technology (VCQ), A-1090 Vienna, Austria}
\affiliation{Department of Microtechnology and Nanoscience (MC2), Chalmers University of Technology, S-412 96 Gothenburg, Sweden}
\author{M. Aspelmeyer}
\affiliation{University of Vienna, Faculty of Physics, Vienna Center for Quantum Science and Technology (VCQ), A-1090 Vienna, Austria}
\affiliation{Institute for Quantum Optics and Quantum Information (IQOQI), Austrian Academy of Sciences, A-1090 Vienna, Austria}

\date{\today}

\begin{abstract}
We provide more information on: (i)The field distribution of the magnetic trap, the resulting trap frequencies and the derivation of the coupling strength. (ii)Effects of current fluctuations and drifts on the measured spectral density. (iii)Limits to the force sensitivity due to measurement noise and current drifts. (iv)Feedback control of the levitated particle. (v)The vibration isolation system. (vi)Possible contributions to the damping. (vii)Calibration of the SQUID signal. (viii)Optimizations and requirements for ground state cooling.
\end{abstract}

\maketitle

\section{Derivation of the magnetic trap's properties}

\subsection{Trapping field}
\begin{figure}
\includegraphics[width=\columnwidth]{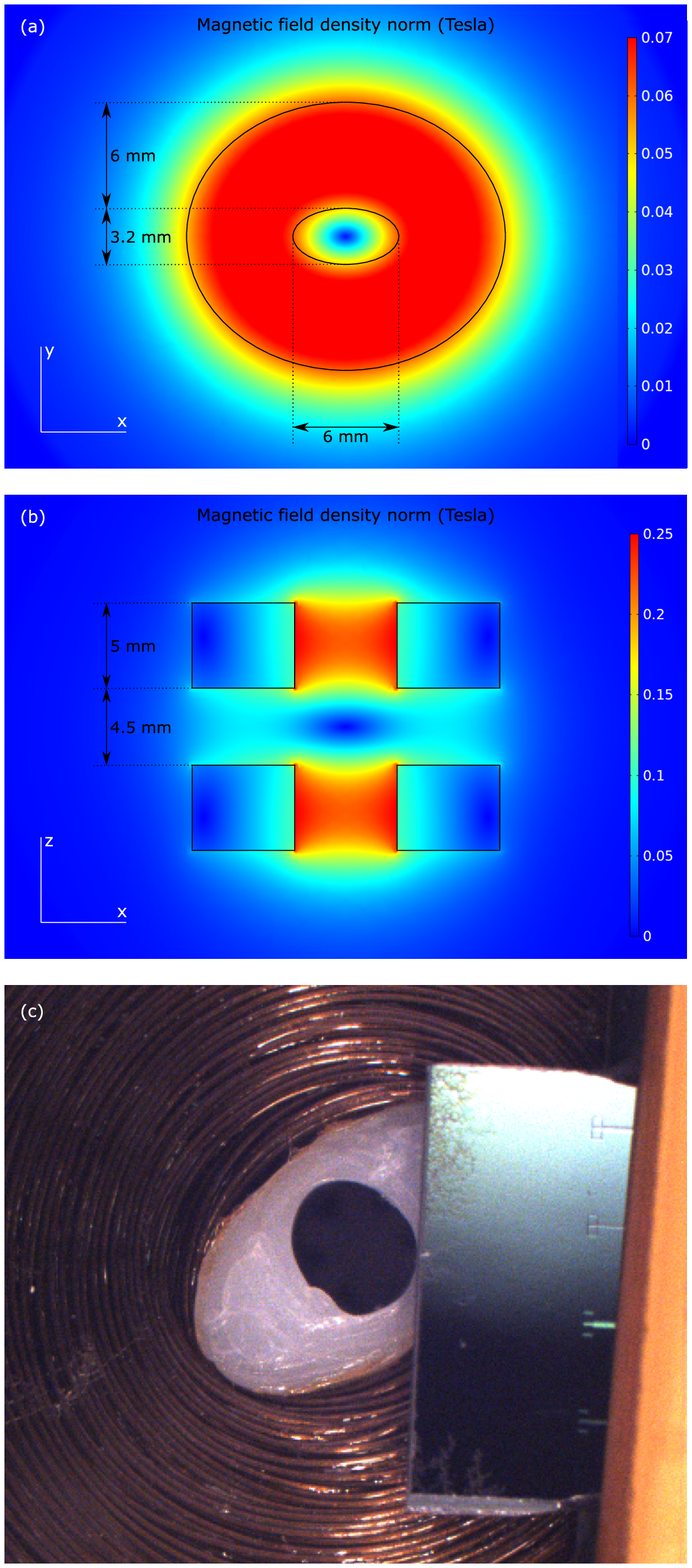}
\caption{\label{figS1} (a,b)Numerical simulations of the magnetic field density created by coaxial elliptical coils with counter-propagating currents. (c)Photograph of the top surface of a single trap coil. On the right side the wafer with the pickup loops is visible (not in its final position).}
\end{figure}

We performed numerical simulations in \textsc{comsol multiphysics} to predict the field configuration generated by the arrangement of our trapping coils. The design of the coils and the simulated magnetic field density are shown in Fig.~\ref{figS1}a,b. 
The field gradients for a trap current of \qty{2.5}{A} are obtained as $(b_{x},b_{y},b_{z})=(\num{57},\num{90},\num{147})\unit{T/m}$, which would correspond to frequencies of $(\num{92},\num{144},\num{236})\unit{Hz}$. Comparing this to the measured values of $(\num{109},\num{127},\num{236})\unit{Hz}$ we can see that the agreement between the numerical and measured z-mode frequency is excellent, but that the simulations predict a much stronger split for the x- and y-mode than what we see in the experiment. This is likely due to the fact that the spooling process resulted in coils that were more circular than the design (cf.\ Fig.~\ref{figS1}c). For a perfectly circular coil the x- and y-mode frequencies are degenerate \cite{Hofer2019}, so a smaller frequency split is to be expected.
Close to the center the field is very well approximated by a quadrupole field: Within a cubic volume of \qtyproduct{200x200x200}{\micro\metre} the maximum relative root-mean-square deviation, defined as $\vert\mathbf{B_{sim}}-(b_{x} x,b_{y} y, b_{z} z)\vert/\vert\mathbf{B_{sim}}\vert$, is less than \qty{0.1}{\percent}.
 
\subsection{Mechanical frequencies}
In this section we build on the calculations in \cite{Hofer2019} to analytically calculate the magnetic field distribution for a superconducting sphere in an axially-asymmetric magnetic quadrupole field.
We consider a superconducting sphere at the origin of the coordinate system and displaced from the center the quadrupole field by $\mathbf{x_{0}}=(x_{0}, y_{0}, z_{0})$. The applied field then reads 
\[
\mathbf{B_{0}} = \left(b_x(x+x_{0}),b_y(y+y_{0}),b_z(z+z_{0})\right).
\]
As our sphere's radius $R$ is much larger than the expected penetration depth $\lambda$ \cite{Livingston1963}, we can use the approximation $\lambda/R\rightarrow 0$, i.e. we assume $\mathbf{B}=0$ inside the sphere and $B_r=0$ on the sphere's surface \cite{Hofer2019}, where $B_r$ denotes the radial component of the field. As there are no currents outside the sphere, there exists a scalar potential $\Phi$ s.t. $\mathbf{B}=\mathbf{B_{0}}-\nabla\Phi$ and
\[
\Phi=\sum_{n=0}^{\infty}r^{-(n+1)}\sum_{m=-n}^n a_{n,m} Y_n^m,
\]
where $r=\sqrt{x^2+y^2+z^2}$ and the $Y_n^m$ are spherical harmonics. The coefficients $a_{n,m}$ are determined by the boundary condition $B_{0,r}=(\nabla\Phi)_{r}$ on the sphere's surface as
\begin{eqnarray*}
a_{1,0} =& -&b_{z}\sqrt{\pi/3}~R^{3} z_{0},\\
a_{1,-1} =& -&\sqrt{\pi/6}~R^{3} (b_{x} x_{0} + i b_{y} y_{0}),\\
a_{1,1} =& &\sqrt{\pi/6}~R^{3} (b_{x} x_{0} - i b_{y} y_{0}),\\
a_{2,0} =& -&b_{z}\sqrt{4\pi/45}~R^{5},\\
a_{2,-2} =& &(b_{y}-b_{x}) \sqrt{2\pi/135}~R^{5},\\
a_{2,2} =& &a_{2,-2},
\end{eqnarray*}
all other coefficients are zero.
From this field distribution we obtain the force on the sphere as
\[
\mathbf{F}=-\frac{3V}{2\mu_0}\left(b_{x}^{2} x_{0},b_{y}^{2} y_{0},b_{z}^{2} z_{0}\right),
\]
where $V$ is the volume of the sphere. The magnetic field thus creates a harmonic trapping potential for a superconducting sphere, with trapping frequencies described by 
\[
f_i = \sqrt{\frac{3}{8\pi^2\mu_0\rho}}|b_i|.
\]

\subsection{Coupling strength}
\begin{figure}
\includegraphics[width=\columnwidth]{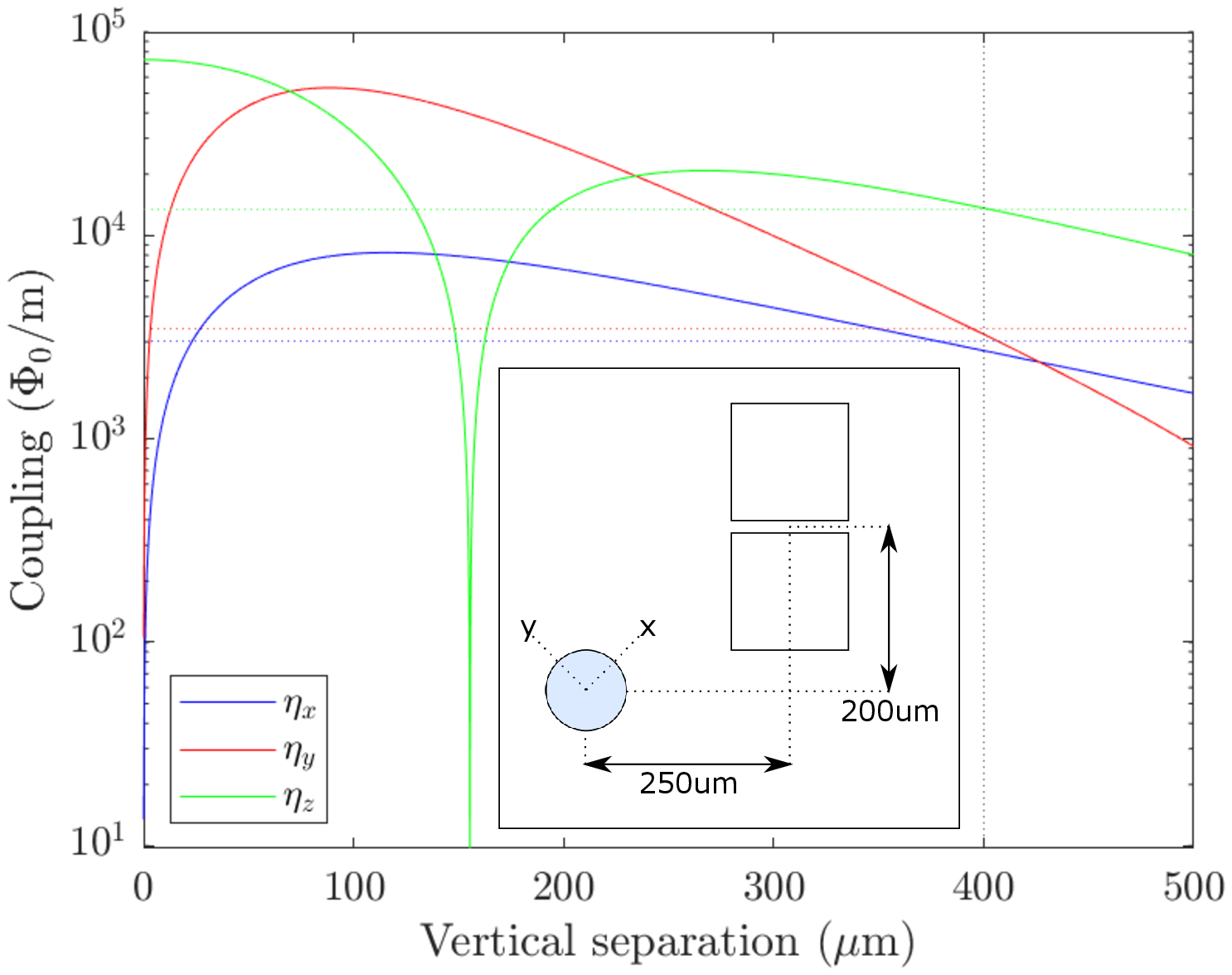}
\caption{\label{figS2} Dependence of the coupling strength on the vertical separation between trap center and pickup loop, when the separation along x and y is as shown in the inset - these are the values we estimate for our setup. The vertical separation in our setup is approximately \qty{400}{\micro\metre}. The dotted horizontal lines correspond to the measured coupling strengths.}
\end{figure}

The linear coupling strength with respect to a pickup loop that is described by a closed path $\gamma$ is defined as $\nu_{i}=\partial_{i}\Phi_{\gamma}$, where $\Phi_{\gamma}$ is the flux through the area enclosed by the pickup loop and $i\in\{x_{0},y_{0},z_{0}\}$.
We can express this as
\begin{equation}
\label{eqS1}
\nu_i=\int_\gamma\mathrm{d}\mathbf{\gamma}~\partial_i\mathbf{A},
\end{equation}
where $\mathbf{A}$ denotes the magnetic vector potential, defined by $\nabla\times\mathbf{A}=\mathbf{B}$ and $\nabla\mathbf{A}=0$. While the resulting expressions are generally bulky, they can be easily evaluated using a computer algebra system, we use \textsc{Wolfram Mathematica}.
We measure the current that is induced in the pickup loop using a SQUID current sensor, i.e. the pickup loop is connected to an input coil on the same wafer as the SQUID and inductively coupled to the SQUID.
The coupling strength with respect to the SQUID is thus obtained as $\eta_i=\nu_i M/L$, where $L$ is the inductance of the circuit formed by pickup loop, input coil of the current sensor, and the connecting wires, while $M$ is the mutual inductance between input coil and SQUID. In our case $M/L\approx\num{0.02}$.  
In Fig.~\ref{figS2} we plot the dependence of the coupling strength on the vertical separation between trap center and pickup loop, when the x- and y-position of the pickup loop relative to the trap center (as shown in the inset) is approximately the same as what we estimate for our setup (from pictures taken at room temperature, cf.\ Fig.~\ref{figS1}c). The horizontal dotted lines correspond to the measured coupling strengths, while the vertical dotted line corresponds to the vertical separation extracted from camera images (approx. \qty{400}{\micro\metre}) of the levitated sphere.

\subsection{Trap displacement due to gravity}
The rest position of the sphere is slightly displaced from the field minimum due to gravity. In our case the setup is aligned such that the direction of gravity coincides with the z-axis. The expected displacement is thus 
\[
z_{g} = -g/(2\,\pi\,f_{z})^2,
\]
where $g\approx \qty{9.81}{\metre\per\second^2}$. In general this displacement needs to be taken into account when calculating the coupling strength, but for larger trapping frequencies it becomes negligible. We usually operate at $f_z>\qty{200}{Hz}$, for which $z_{g}<\qty{6}{\micro\metre}$.

\section{Current fluctuations}
\begin{figure}[ht!]
\centering
\includegraphics[width=\linewidth]{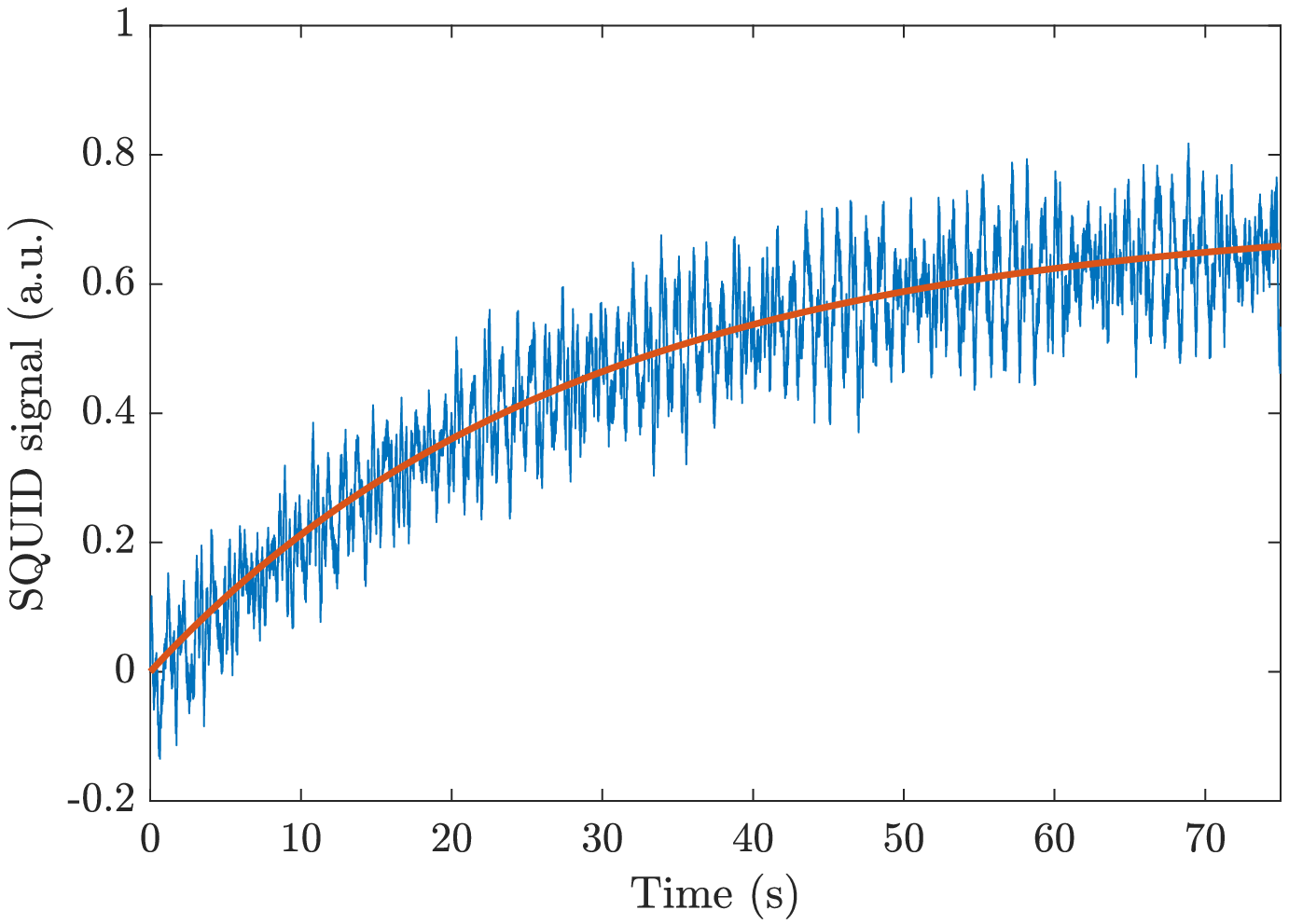}
\caption{\label{figS3} Step response of the filter measured using the SQUID. The solid line is a fit of the filter's step response $1-\mathrm{exp}(-R_{C}/L_{C}\, t)$.} 
\end{figure}

Since the sphere's motional frequencies depend linearly on the trap current, current fluctuations result in frequency fluctuations, and thus can cause heating as well as broadening of the mechanical spectral peaks. We stabilize the trap current both passively and actively, as described in the following.
We implement a first-order low-pass RL-filter by adding a short copper wire (resistance $\mathrm{R_{C}}$) in parallel to the trap coils (inductance $\mathrm{L_{C}}$). We characterize the filter's cut-off frequency by measuring its step response in an empty trap (Fig.~\ref{figS3}). The cutoff frequency is $R_{C}/L_{C}\approx \qty{0.036}{s^{-1}}$, which results in an attenuation of approximately \qty{-180}{dB} at \qty{200}{Hz}, while the added thermal current noise from the copper wire is only on the order of \qty{1e-11}{A/\sqrt{Hz}}.

\begin{figure}[ht!]
\centering
\includegraphics[width=\linewidth]{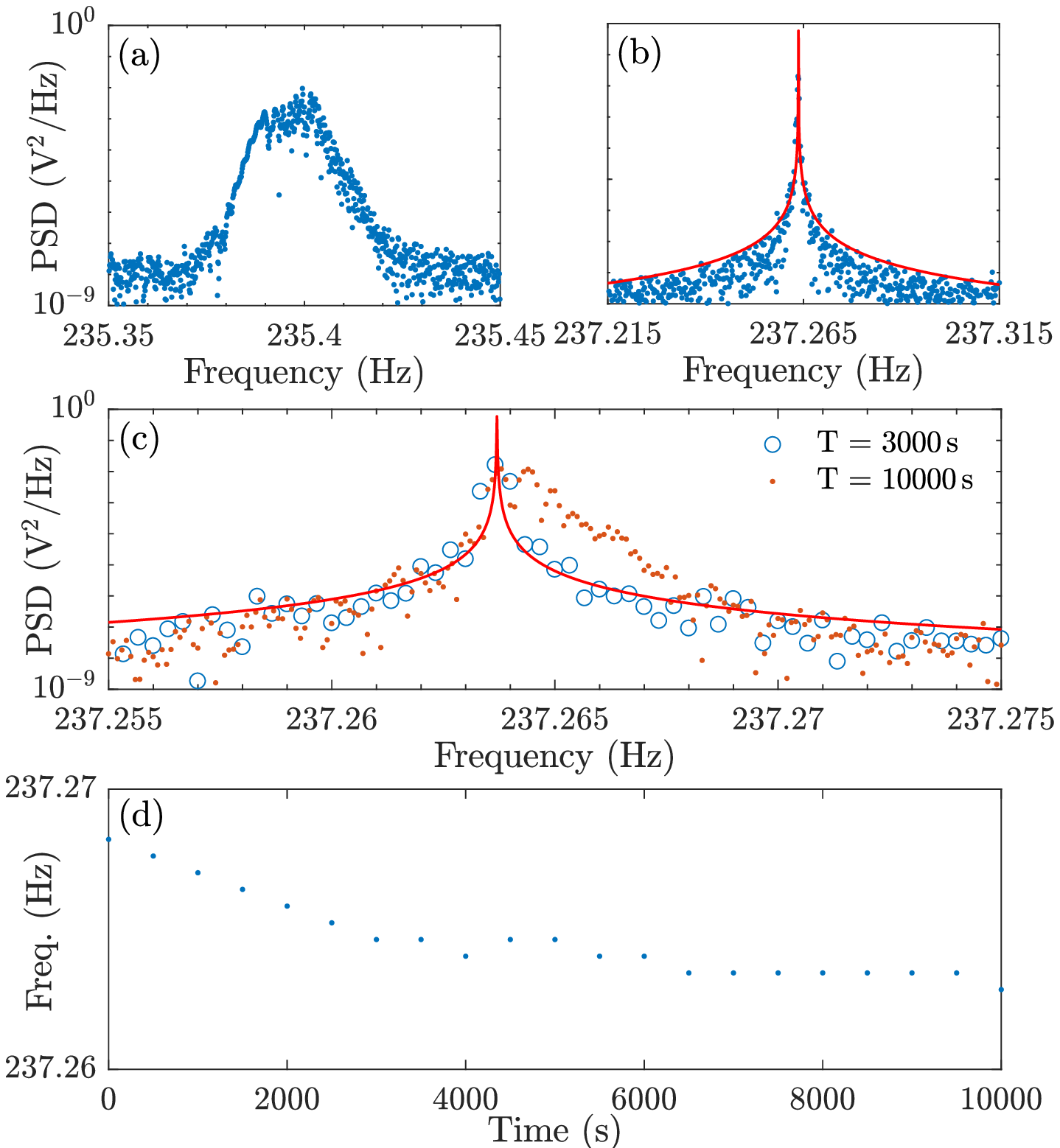}
\caption{\label{figS4} (a) Typical motional resonance without active current stabilization. (b) Typical motional resonance with active current stabilization. (c) Comparison of PSDs for different measurement lengths T with trap current feedback on. For short T the linewidth is determined by the spectral resolution 1/T and spectral leakage, while for longer T there is still linewidth broadening due to drifts in frequency. The solid line in (b) and (c) is a Lorentzian with the linewidth set by a ringdown measurement. (d) The resonance frequency drifts by around \qty{5}{mHz} over three hours.} 
\end{figure}

We actively stabilize the filtered trap current by monitoring the mean current source output using a digital ammeter and updating the current source output based on an estimation of the trap current.
This lessens current drifts and the associated resonance broadening, as seen by comparing Fig.~\ref{figS4}(a) with (b).
Even with active current stabilization we observe drifts of the particle's resonance frequency shown in Fig.~\ref{figS4}(d), which we attribute to low-frequency noise in the ammeter reading.

\section{Sensing}
\begin{figure}[ht!]
\centering
\includegraphics[width=\linewidth]{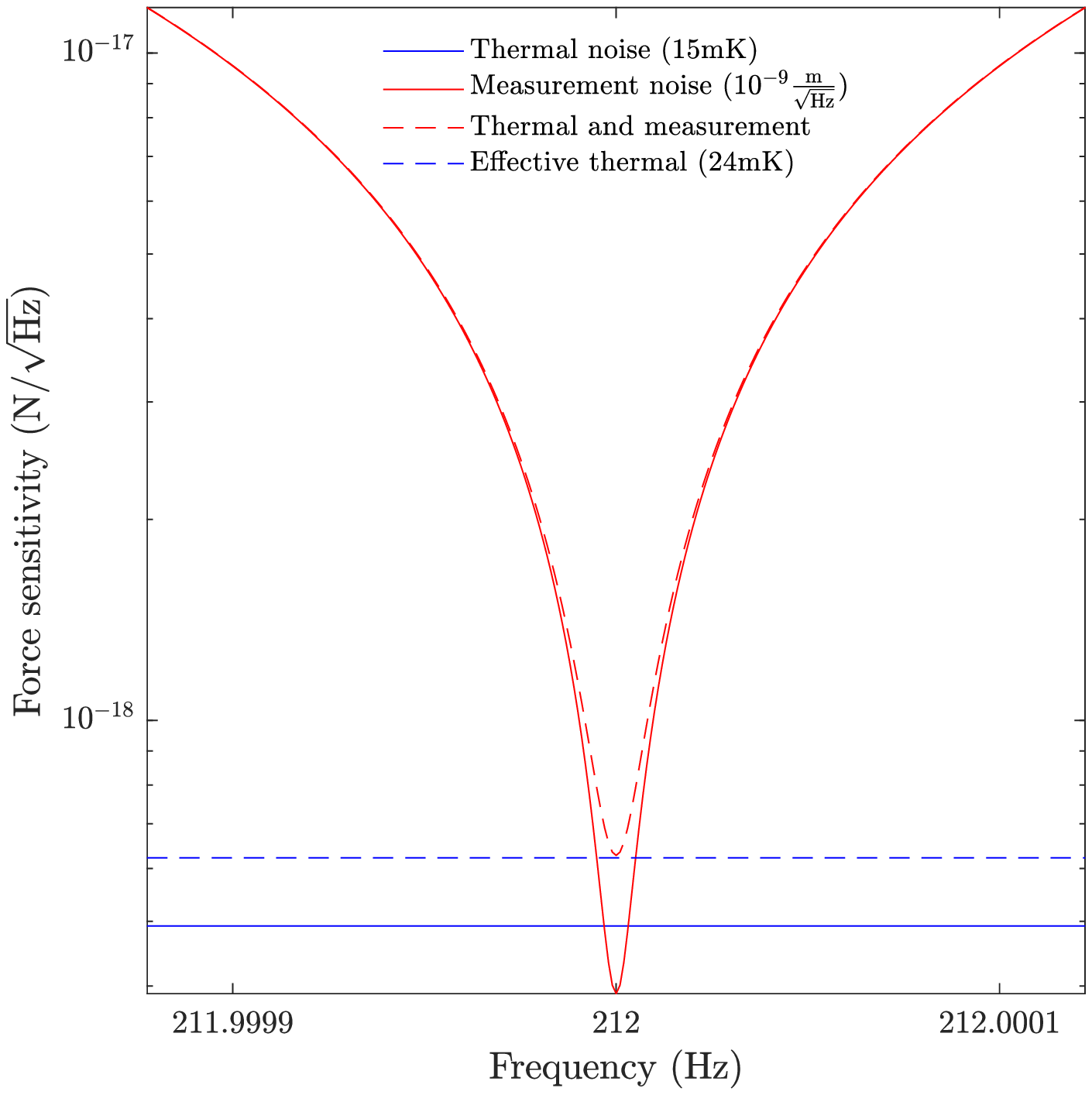}
\caption{\label{figS9} (a)Force sensitivity limits due to thermal noise and measurement noise. The parameters used for the plot are $Q=\num{2.6e7}$ and $f_{0}=\qty{212}{Hz}$.} 
\end{figure}

Taking into account measurement noise, we can calculate the force sensitivity as
\[
\sqrt{S_{FF}}=\sqrt{4k_{B}T_{0}m\gamma + |\chi|^{-2}S_{nn}},
\]
where $\sqrt{S_{nn}}\approx \qty{1e-9}{m/\sqrt{Hz}}$ is the measurement noise and $\chi(\omega)=1/[m(\omega_0^2-\omega^2-i\gamma\omega)]$ denotes the mechanical susceptibility. In our case, using $Q=\num{2.6e7}$ at \qty{212}{Hz} and assuming $T_{0}=\qty{15}{mK}$, this would result in a force sensitivity of $\sqrt{S_{FF}}=\qty{6.3e-19}{N/\sqrt{Hz}}$ on resonance, close to the thermal limit of \qty{4.9e-19}{N/\sqrt{Hz}}, and a noise equivalent temperature of $T_{n}=T_{0}+|\chi(\omega_{0})|^{-2}S_{nn}/(4k_{B}m\gamma)=\qty{24}{mK}$ (cf. Fig.~\ref{figS9}). 
However, due to the stochastic frequency fluctuations described in the last section the susceptibility is stochastic as well. We briefly sketch how this reduces our sensitivity: Note that a measurement with frequency drifts can be approximated as a series of shorter measurements at different, but constant, resonance frequencies. For a finite measurement duration $T$ the measured force power spectral density at resonance is approximately
\[
4k_{B}T_{0}m\gamma + S_{nn}\frac{\Delta f}{\int_{\Delta f}df|\chi(f)|^{2}},
\]
where the integration runs over the bin width $\Delta f = \frac{1}{T}$ of the bin containing the resonance. Our force sensitivity thus becomes $\sqrt{S_{FF}}=\sqrt{4k_{B}T_{0}m\gamma + |\chi|_{avg}^{-2}S_{nn}}$ with $|\chi|_{avg}^{2}=\frac{1}{\Delta f}\int_{\Delta f}df|\chi(f)|^{2}$. For $\Delta f>\frac{\gamma}{2\pi}$, we get $\frac{|\chi(f_{0})|^{2}}{|\chi|_{avg}^{2}}\propto\frac{\Delta f}{\gamma}$. In our case this boosts the measurement noise above the thermal noise and, for a typical measurement, our force sensitivity worsens by at least an order of magnitude, corresponding to an noise equivalent temperature of at least \qty{2.5}{K}.
Both improving the coupling, which corresponds to an effective decrease of the measurement noise, and improving the stability of the resonance frequency will allow us to perform measurements limited by thermal noise. We show below that we can increase the coupling by almost four orders of magnitude. Regarding the latter, we note that persistent current coils have reached a relative current stability better than \num{2e-11}/\unit{\per h} \cite{Van1999}.

\section{Feedback}
We can apply a magnetic feedback force on the levitated sphere by processing the SQUID signal and applying a feedback current to a small coil with approximately 20 windings positioned approximately \qty{1}{mm} below the trap center. The feedback current generates an additional magnetic field, thereby shifting the field minimum of the trapping field and enabling us to apply direct feedback. The gradient of the trapping field is changed as well, such that we also have the possibility to apply parametric feedback at twice the particle's resonance frequencies. Since the COM modes are separated in frequency, we can apply feedback to all modes simultaneously. 
When we apply direct feedback, we pass the SQUID signal to a FPGA (STEMlab 125-14), which, for each of the mechanical frequencies, applies a bandpass-filter, gain and phase shift \cite{Neuhaus2017}. The resulting signals are then recombined on the FPGA and fed back to the feedback coil. When we apply parametric feedback at twice the mechanical frequencies we additionally use a phase-locked loop and a clock divider on a second FPGA.
As described in the last section, our current noise floor corresponds to a noise equivalent temperature that is higher than the base temperature of our dilution refrigerator. This noise equivalent temperature also sets the limit for feedback cooling. Due to the large ambient noise reduction provided by our vibration isolation, magnetic shielding and trap current filter, the equilibrium temperature of the particle is below the noise equivalent temperature, meaning that even without applied feedback the particle's motion undergoes exponential decay until it is not detectable anymore. We thus cannot cool the particle's modes below their equilibrium temperature and use feedback currently only as a way to quickly adjust the amplitudes and prepare other measurements.

\section{Characterization of the vibration isolation system}
\begin{figure}[ht!]
\centering
\includegraphics[width=\linewidth]{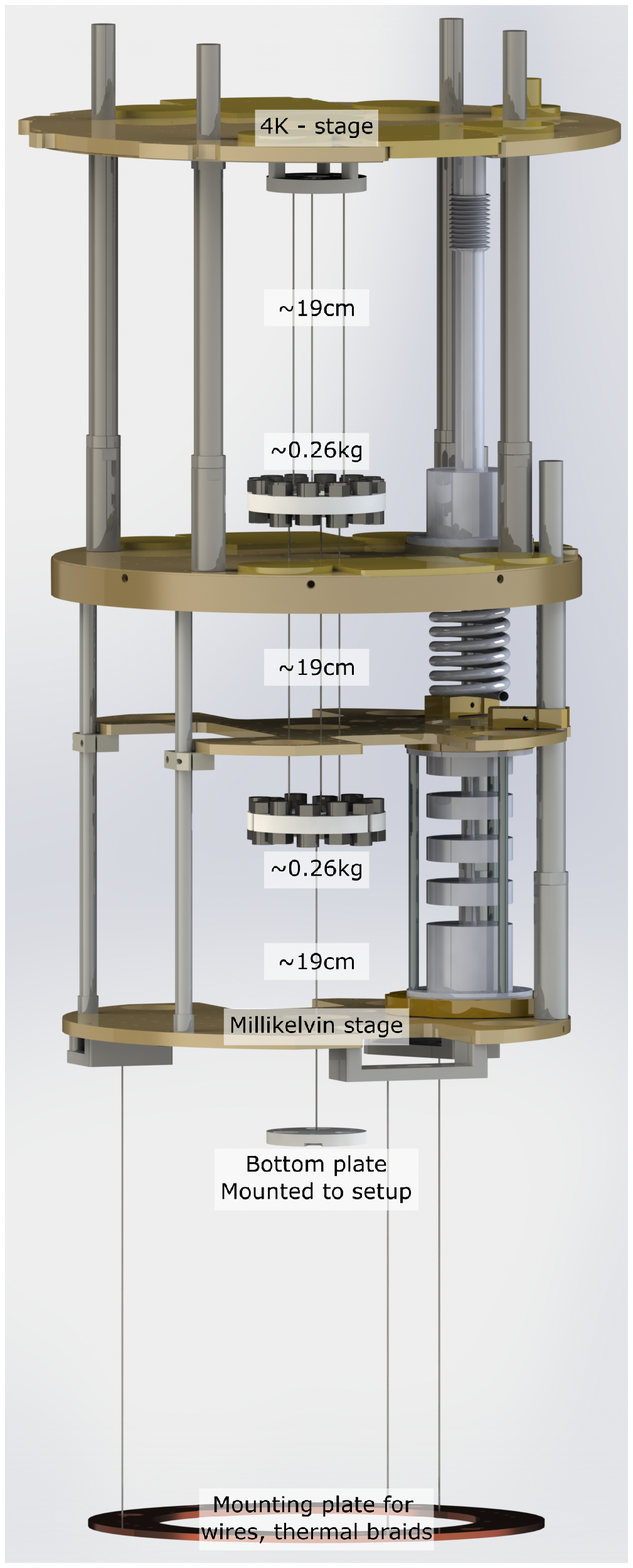}
\caption{\label{figS5}Model of the vibration isolation system.} 
\end{figure}

\begin{figure}[ht!]
\centering
\includegraphics[width=\linewidth]{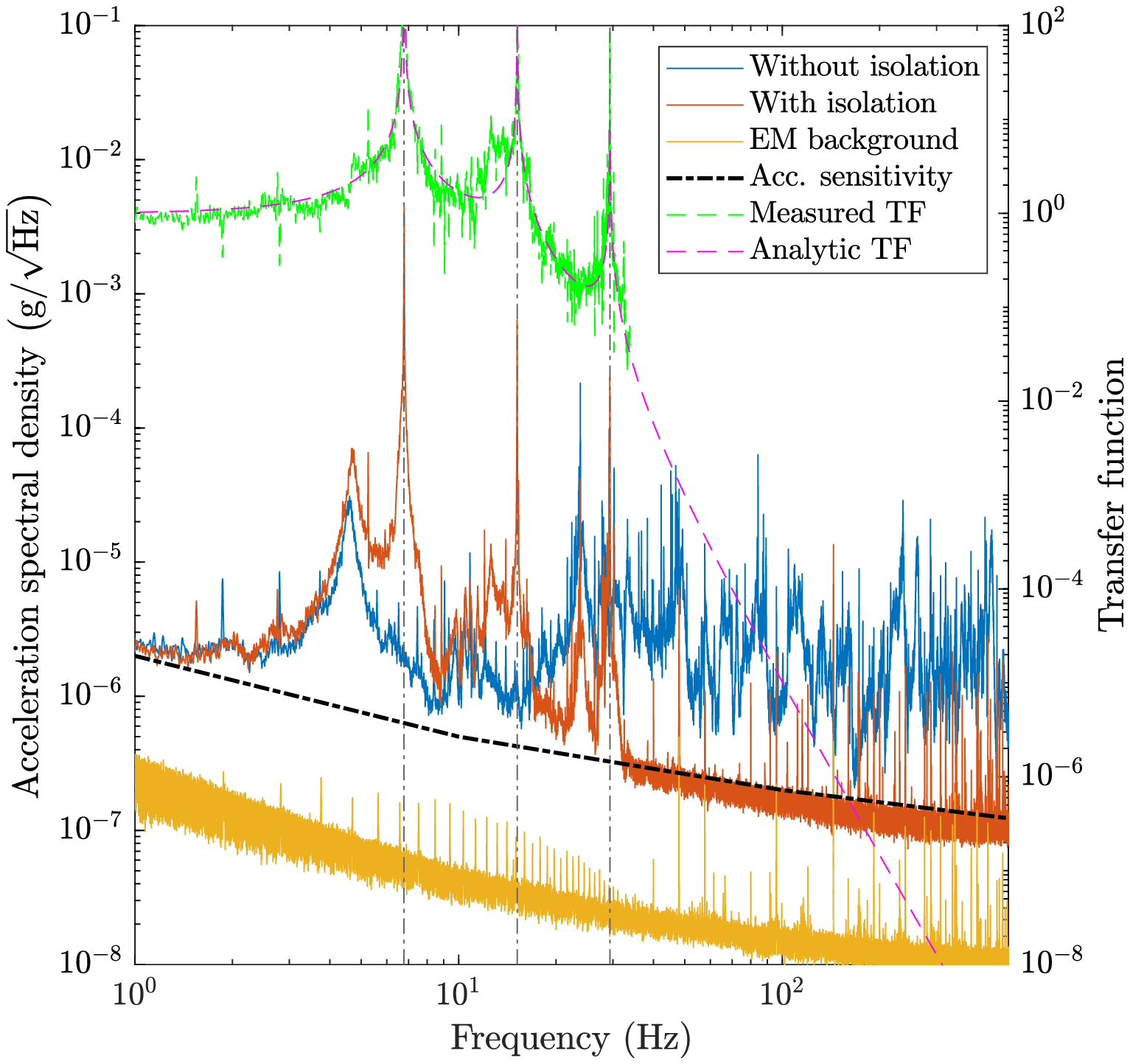}
\caption{\label{figS6}Characterization of the vibration isolation. The dash-dotted black line is the sensitivity of the accelerometer, as specified by the manufacturer. The solid orange and blue lines correspond to the measured acceleration with and without the vibration isolation, respectively, while the solid yellow line is data recorded with a \qty{1}{M\ohm} resistor connected to the SMA port inside the fridge. The dashed pink line is the analytic transfer function (Eq.~\ref{eqS2}), while the dashed green line is the measured transfer function, i.e. the ratio of the measured accelerations. The horizontal lines correspond to the normal mode resonance frequencies.} 
\end{figure}

Our vibration isolation system consists of several cylindrical plates connected by straight wires, with the top plate mounted to the 4K-platform of the dilution refrigerator and the bottom plate mounted to the aluminum can containing the magnetic trap. 
Electrical and thermal connections going to the setup are loosely coiled up and guided to the setup via a separate vibration isolation stage hung from the Mixing Chamber platform (cf. Fig.~\ref{figS5}). 
We first characterized a single stage at room temperature, by exciting it mechanically and recording its motion with a camera. We find that the horizontal and vertical resonance frequencies are $f_{h}=\qty{1.1}{Hz}$ and $f_{v}=\qty{18.7}{Hz}$, respectively, while the librational resonance frequencies are $f_{l,1}=\qty{0.6}{Hz}$ (around vertical axis), $f_{l,2}=\qty{8.7}{Hz}$ and $f_{l,3}=\qty{9.0}{Hz}$ (around horizontal axes). We are thus limited by vertical vibrations, which in general couple to the other degrees of freedom as well \cite{Aldcroft1992}.
We then assembled the setup as shown in Fig.~\ref{figS5}, with the bottom plate mounted to an accelerometer (instead of the setup) used to measure acceleration along the vertical axis. The mass of the bottom plate (including the accelerometer) was \qty{0.29}{kg} in this assembly. The intermediate stages are each supported by three wires of equal length, while the bottom stage is supported by a single wire. The wires are made from type 304 stainless steel with a diameter of $D=\qty{38}{\micro m}$ and a yield load of \qty{0.37}{kg}, as specified by the manufacturer (Fort Wayne Metals). The wires are connected to the stages with clamping connections. 
The vertical spring constant for the i-th stage is given by $k_{v,i}=(N_{i} Y D^2 \pi)/(4 L_{i})$, s.t. $f_{v,i}=1/(2\pi)\sqrt{k_{v,i}/m_{i}}$ is the resonance frequency of a single stage. Here $Y$ is the Young's modulus, $N_{i}$ the number of supporting wires, $L_{i}$ the wire length and $m_{i}$ the mass of the stage. The normal mode frequencies $f_{n,i}$ of the assembled system are obtained by solving the characteristic equation of the system \cite{Aldcroft1992}, resulting in the transfer function
\begin{equation}
\label{eqS2}
\prod_{i=1}^{3}\frac{f_{v,i}^{2}}{\vert f_{n,i}^{2}-f^{2}\vert}.
\end{equation}
In Fig.~\ref{figS6} we compare the vibrations measured without vibration isolation (with the accelerometer mounted to the Mixing Chamber stage of the dilution fridge) and with vibration isolation. 
The highest normal mode frequency is approximately \qty{30}{Hz}, above which the measured accelerations quickly drop below the sensitivity of the accelerometer, when it is mounted to the vibration isolation system. At lower frequencies the measured transfer function corresponds very well to the analytical values and we use the analytic expression to extrapolate the attenuation at higher frequencies, resulting in an attenuation of \num{1e-5} at \qty{100}{Hz} and \num{1.5e-7} at \qty{200}{Hz}. 
The peaks at higher frequencies correspond to the electromagnetic background in the lab, which is verified by an independent measurement in which the accelerometer is replaced by an \qty{1}{M\ohm} resistor inside the fridge. 
This measurement reproduces the peaks at precisely the same frequencies as in the accelerometer measurement, although with a lower magnitude - this we attribute to the additional (unshielded) wiring going from the SMA port to the accelerometer, which can act as an antenna.
In an initial approach we used different masses and wire lengths, such that the attenuation was approximately \num{5} orders of magnitude at \qty{200}{Hz}.

\section{Quality factor}
In this section we estimate contributions to the dissipation due to gas damping and eddy currents. We also consider dissipation in the SQUID and flux creep in the pickup loop, which we rule out experimentally. 

Collisions with background gas result in an additional damping of \cite{Hinkle1990}
\[
\gamma_{P}=\beta \frac{P}{\rho R \bar{v_{th}}},
\]
where $\beta\approx 1.8$ and $\bar{v_{th}}$ is the mean thermal velocity of the molecules. 
We do not have a direct measure of the pressure at the trap location, but we can use the measured pressure at the room temperature side of the vacuum can, $P\approx\qty{1e-6}{mbar}$, to set an upper limit for the expected damping. Assuming the background gas consists mostly of Helium, we get $\gamma_{P}\approx \qty{3e-7}{s^{-1}}$, two orders of magnitude below the measured damping rates.

We avoid eddy current damping by not placing any normal conductors within the inner shield. Eddy currents can still be induced in conductors outside the shield, either via openings in the shield or via the pickup circuit, part of which is placed outside the shield. In the first case, we can estimate possible eddy current losses by imagining a conductive loop with resistance $R_{o}$ and inductance $L_{o}$ placed around the openings in the shield. The dissipation in mode $i\in\{x_{0},y_{0},z_{0}\}$ then takes a maximum value for $R_{o}=2\pi f_{i}L_{o}$, corresponding to a damping 
\[
\gamma_{i}=\frac{(\partial_{i}\phi)^2}{2\pi m f_{i} L_{o}}.
\]
Here $\phi$ denotes the flux in the loop induced by the oscillation of the sphere, which can be calculated from Eq.~\ref{eqS1}. Using the dimensions of our windows and estimating $L_{o}>\qty{10}{nH}$ for a similarly sized conductive loop, we get $\gamma_{i}<\qty{1e-9}{s^{-1}}$ for all modes.

We now consider eddy current damping mediated by the pickup circuit or damping due to flux creep in the pickup circuit, as well as dissipation in the SQUID due to a real part in the SQUID input impedance \cite{Martinis1985}. This implies $\gamma_{i}\propto\nu_{i}^{2}$, while in the experiment the damping is approximately equal for all three COM modes. In addition, during initial measurements the pickup loop was positioned farther from the trap center and coupling strengths were approximately one order of magnitude lower than the ones reported here, but we did not see any change in the quality factor. 

\section{Calibration}
\begin{figure}[ht!]
\centering
\includegraphics[width=\linewidth]{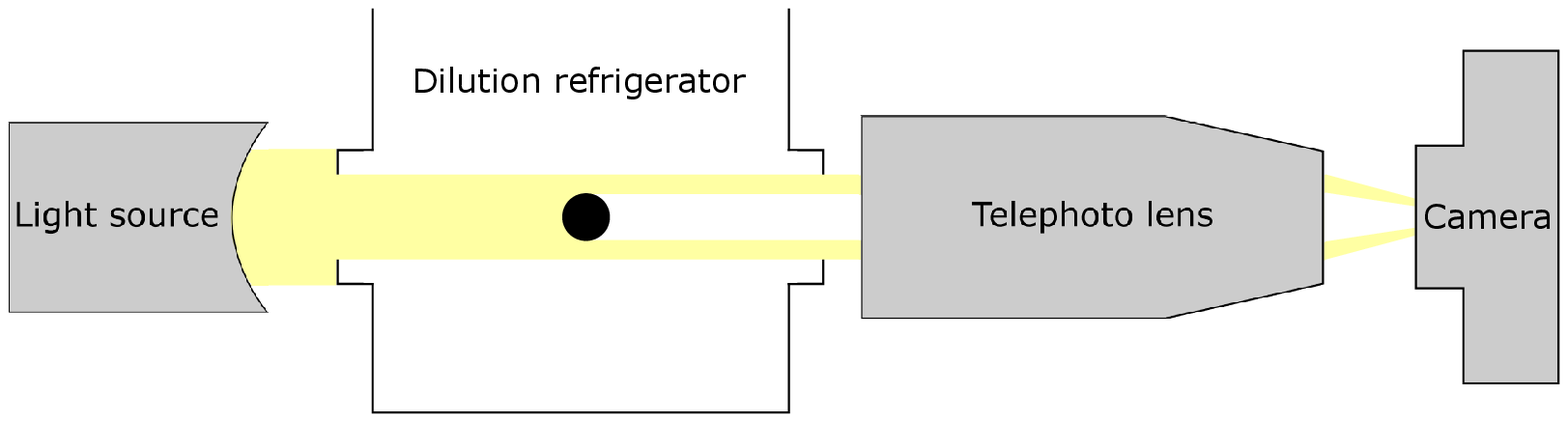}
\caption{\label{figS7}Sketch of the optical setup.} 
\end{figure}

Our optical readout scheme is shown schematically in Fig.~\ref{figS7}. 
The light source (Leica CLS150) produces a beam of light, which is further collimated by small windows in the cryostat and the magnetic shields. After passing through the cryostat, the light is focused by a teleobjective into the camera (DFK 33UX287). Any object in the light path thus appears as a shadow on a bright background, corresponding to the projection of the object onto the image sensor.
We use a frame rate of 596fps and 0.5ms exposure time to record the videos and then process them as follows.
Each frame is first converted to grayscale, smoothed with Gaussian blurring, thresholded and then analyzed using the particle tracking software \textsc{Trackpy} \cite{Crocker1996}. This results in datasets for the horizontal and vertical (as defined by the camera's orientation) position of the sphere. The camera is aligned such that the vertical axis is along the trap coil axis and the direction of the particle's z-motion, while the particle's motion along x and y is projected onto the camera's horizontal axis. The trap coils are aligned such that the elliptic axes are at around $\qty{45}{\degree}$ to the optical axis, so both the x and y motion can be imaged. To get separate trajectories for each mode we implement a digital bandpass filter around the respective resonance frequency. Calibration of the sphere's displacement relies on knowledge of the optical system's magnification, which we determine using knowledge of the sphere's size (\qty{100\pm6}{\micro m}, as specified by the manufacturer). Including the uncertainty from the alignment of the trap coils with respect to the optical axis in addition to the \qty{\pm6}{\percent} uncertainty from the size of the sphere, we estimate a total uncertainty of \qty{\pm13}{\percent} for the calibrated COM displacement.

The light source is turned on only for calibration measurements, as the illumination heats the particle and limits the levitation time to approximately one hour. Further, optical band-pass filters (approximately \qty{400}{nm} -- \qty{650}{nm}) on the millikelvin stage reduce ambient light reaching the particle and heating it. Without the filters (and the light source turned off) levitation times are limited to around four hours, while we have not yet found a limit (\textgreater \qty{30}{h}) using the filters.

\section{Optimizations and ground state cooling}
In this section we provide more detail on how to implement the improvements that are necessary to perform feedback cooling to the ground state. We will consider only cooling of the z-mode, cooling of the other modes and 3D-cooling can be characterized and optimized in an analogous manner. We first show that the coupling strength can be increased by almost four orders of magnitude, s.t. the measurement noise (in units of \unit{m/\sqrt{Hz}}) with a state-of-the-art SQUID decreases to $\qty{1e-15}{m/\sqrt{Hz}}$. We then take into account the back action from the SQUID due to the circulating current and determine the standard quantum limit (SQL) of our system. We proceed to evaluate the requirements for ground state cooling as well as contributions to heating from vibrations and frequency fluctuations. 

\subsection{Coupling strength}
\begin{figure}
\includegraphics[width=\columnwidth]{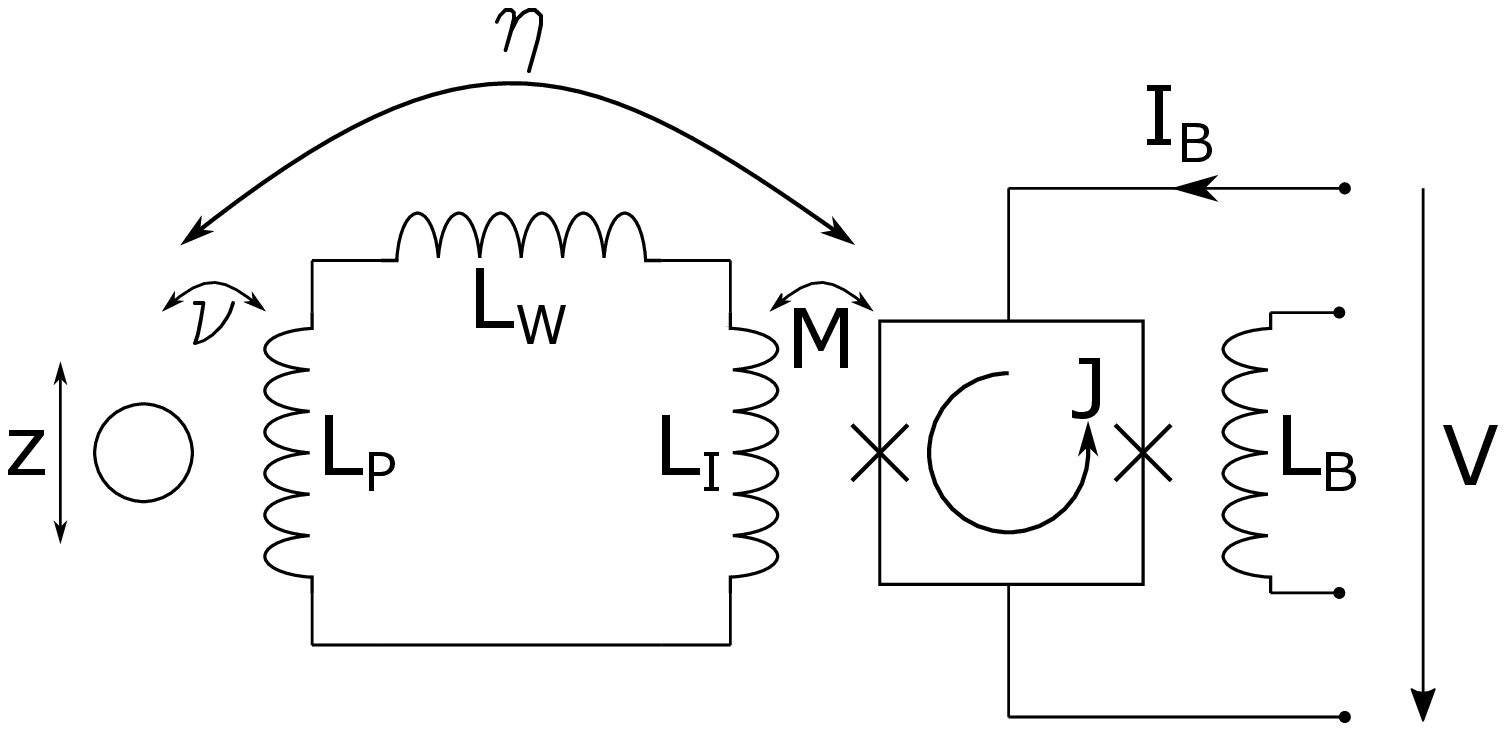}
\caption{\label{figS8} Schematic of the readout circuit. The working point of the SQUID is set by the bias current $I_{B}$ and the bias flux coupled into the SQUID via $L_{B}$. For appropriate bias values and small changes in flux (in FLL mode this is assured by applying feedback over $L_{B}$), the voltage drop $V$ is proportional to the flux in this SQUID loop.}
\end{figure}

We can increase the coupling strength by better positioning of the pickup loop relative to the superconducting sphere and by increasing the number of turns of the pickup loop, thus matching its inductance to that of the input coil. For the calculation, we will assume a planar pickup loop with circular windings, positioned coaxial with the z-axis. Then Eq.~\ref{eqS1} leads to 
\begin{eqnarray}
\label{eqS3}
\nu_{1}=&&\pi b_{z} R_{P}^{2}\left(\frac{R^2}{R_{P}^{2}+Z_{P}^{2}}\right)^{\frac{3}{2}}\nonumber\\
&&\times\left(1-\frac{R^2}{R_{P}^{2}+Z_{P}^{2}}\left(1-\frac{5 Z_{P}^{2}}{R_{P}^{2}+Z_{P}^{2}}\right)\right)
\end{eqnarray}
for the coupling strength w.r.t. a single turn of the pickup loop with radius $R_P$ and z-position $Z_P$. The coupling strength to a multi-turn pickup loop can be written as a sum of terms of this form, $\nu=\sum_{i}\nu_{1}(R_{P,i},Z_{P,i})$, and the coupling strength to the SQUID (cf. Fig.~\ref{figS8}) is obtained as
\[
\eta = \nu\frac{k\sqrt{L_{I}L_{S}}}{L_{P}+L_{I}+L_{W}}.
\]
Here $L_{P}$, $L_{I}$ and $L_{S}$ are the inductances of pickup coil, SQUID input coil and SQUID, respectively. We have used that the the mutual inductance $M$ between input coil and SQUID can be written as $M=k\sqrt{L_{I}L_{S}}$ with $|k|<1$. $L_{W}$ denotes the stray inductance of the pickup circuit and is in our case dominated by the wires connecting the pickup coil to the SQUID, $L_{W}\approx 100nH$. 

A relevant figure of merit for the measurement noise of the SQUID is the so-called energy resolution $S_{EE}=\frac{S_{\phi\phi}}{2L_S}$, with state-of-the-art SQUIDs reaching values on the order of $\hbar$ \cite{Awschalom1988,Carelli1998}. In terms of position resolution this can be written as 
\begin{equation}
\label{eqS4}
S_{nn}=\frac{S_{\phi\phi}}{\eta^2}=\frac{2S_{EE}}{k^{2}}\frac{(L_{P}+L_{I}+L_{W})^2}{\nu^2 L_{I}}. 
\end{equation}
In order to get a realistic estimate for this we will use the SQUID parameters presented in \cite{Carelli1998}, i.e. $L_{S}=\qty{15}{pH}$, $L_{I}=\qty{0.53}{\micro H}$, $M=\qty{2.3}{nH}$ and $\frac{S_{\phi\phi}}{2 k^{2} L_S}=5.5\hbar$. For the pickup coil we assume a wire width of \qty{0.3}{\micro m} and a distance between the wires of \qty{0.45}{\micro m}, which is also the resolution of the input coil of our existing SQUID. We use the modified Wheeler formula \cite{Wheeler1928} to determine $L_{P}$, s.t. we get a fully analytical expression for $S_{nn}$. 
We also restrict us to $Z_{P}\geq R$. Given these parameters, as well as $b_{z}=\qty{147}{T/m}$, we minimize $S_{nn}$ with respect to $R_{P},Z_{P}$ and $N$, where $R_{P}$ now denotes the inner radius of the pickup coil and $N$ is the number of turns. This results in $\eta\approx\qty{5.5e7}{\phi_{0}/\metre}$ and $S_{nn}\approx \qty{1e-15}{m}$ for $Z_{P}=R$, $R_P\ll R$ and $N\approx 87$. 
We note that the scaling of the coupling (Eq.~\ref{eqS3}) means that the measurement noise can be further decreased by working with higher gradients or using larger particles (a scale-independent expression for the coupling is easily obtained by measuring distance in units of $R$, magnetic fields in units of $b_{z} R$ and thus the coupling in units of $b_{z} R^{2}$). 

\subsection{The standard quantum limit}
Equation \ref{eqS4} for the readout signal measured with the SQUID is valid only when the back-action stemming from the circulating SQUID current onto the levitating particle is negligible. This is a reasonable assumption given the weak coupling in our current setup, but for increased coupling strength the back-action should be considered. 
All relevant parameters are stated for the SQUID coupled to the input circuit and can be different than the parameters of the uncoupled SQUID, depending on the capacitive coupling at the Josephson frequency \cite{Martinis1985}. The readout noise is set by the flux noise, while the back-action noise is determined by fluctuations of the circulating current $J$. 
For a shunted SQUID the origin of both flux and current fluctuations is current noise in the shunt resistors and the quantum limit that arises from zero-point-fluctuations is $\sqrt{S_{\phi\phi}S_{JJ}-S_{\phi J}^{2}}\geq\hbar$ \cite{Koch1981,Danilov1983}. 
We will assume in the following that correlations between flux and current noise are negligible, $S_{\phi J}\approx 0$ \cite{Danilov1983,Tesche1979}.

A current $J$ around the SQUID loop will effect a force $-\eta J$ on the levitated sphere and thus the measured displacement PSD of the sphere becomes
\[
S_{zz}^{m} = \frac{S_{\phi\phi}}{\eta^2} + |\chi|^2(S_{FF}^{th}+\eta^{2}S_{JJ}).
\]
The product of measurement noise and back-action force noise thus fulfills the Heisenberg-like inequality $\frac{S_{\phi\phi}}{\eta^2}\,\eta^{2}S_{JJ}\geq\hbar^{2}$ and $S_{zz}^{m}$ is minimized for $\eta^{2}=\frac{1}{|\chi|}\sqrt{\frac{S_{\phi\phi}}{S_{JJ}}}$, such that
\[
S_{zz}^{m} = 2|\chi|\sqrt{S_{\phi\phi}S_{JJ}} + |\chi|^2 S_{FF}^{th}.\]
For $\sqrt{S_{\phi\phi}S_{JJ}}=\hbar$ this corresponds to the standard quantum limit.

\subsection{Feedback cooling}
We now assume we provide direct feedback s.t. the effective damping of the oscillator becomes $\gamma+\Gamma$, where $\Gamma$ is the cold damping added by the feedback \cite{Penny2021} and $\tilde{\chi}(\omega)=1/[m(\omega_0^2-\omega^2-i(\gamma+\Gamma)\omega)]$ is the effective susceptibility. The measurement noise will also enter the feedback system, resulting in an additional force, and the displacement power spectral density becomes 
\[
S_{zz} = |\tilde{\chi}|^2\left(S_{FF}^{th}+\eta^{2}S_{JJ}+(m\omega\Gamma)^2\frac{S_{\phi\phi}}{\eta^{2}}\right).
\]
For $\Gamma=\frac{\eta^{2}}{m\omega_{0}\tilde{L_{S}}}\gg\gamma$ this corresponds a mean phonon number 
\[
\bar{n} = \frac{1}{\hbar\eta^{2}}\left(k_{B}T_{0}m\gamma \tilde{L_{S}}\right)+\frac{1}{2}\left(\frac{\sqrt{S_{\phi\phi}S_{JJ}}}{\hbar}-1\right),
\]
where we have introduced the shorthand $\tilde{L_{S}}=\sqrt{\frac{S_{\phi\phi}}{S_{JJ}}}$. Close to the optimum working point $\tilde{L_{S}}\approx L_{S}$ \cite{Tesche1979}.
Using $\eta=\qty{5.5e7}{\Phi_0/m}$ as well as assuming $T_{0}=\qty{15}{mK}$, $m=\qty{5.6}{\micro g}$ and $L_{S}=\qty{15}{pH}$, we need $\gamma\approx\qty{1e-6}{s^{-1}}$ to keep the left term negligible. Ground state cooling then requires $\sqrt{S_{\phi\phi}S_{JJ}}<3\hbar$. While current noise in SQUIDs is challenging to measure \cite{Martinis1983}, a DC-SQUID with $\sqrt{S_{\phi\phi}S_{JJ}}\approx 10\hbar$ has been reported \cite{Falferi2008}. Further, for $\tilde{L_{S}}\approx L_{S}$ we would expect $\sqrt{S_{\phi\phi}S_{JJ}}\approx 2 S_{EE}$. The lowest energy resolution reported up to date is $S_{EE}\approx 1.7\hbar$ \cite{Awschalom1988}, so ground state cooling by continuous feedback will require a state-of-the-art SQUID.

\subsection{Heating}
We now consider fluctuations of the trap center with spectral density $S_{\epsilon\epsilon}$ as well as fractional fluctuations of the spring constant with spectral density $S_{\delta\delta}$. We note that the heating rate $\Gamma_{\delta}E$ due to frequency fluctuations is proportional to the energy $E$ of the oscillator. 
The average energy of the oscillator is then described by 
\[
\dot{\langle E\rangle}=(\Gamma_{\delta}-\gamma)\langle E\rangle + k_{B}T_{0}\gamma + \dot{Q}_{\epsilon},
\]
where $\dot{Q}_{\epsilon}$ is the heating rate due to fluctuations of the trap center.
For $\Gamma_{\delta}<\gamma$ this results in an effective temperature of 
\[
T_{eff}=\frac{k_{B}T_{0}\gamma + \dot{Q}_{\epsilon}}{k_{B}(\gamma-\Gamma_{\delta})}.
\]
The heating rates are given by \cite{Gehm1998}
\[
\dot{Q}_{\epsilon}=\frac{1}{4}m\omega_{0}^{4}S_{\epsilon\epsilon}(\omega_{0}),
\]
and
\[
\Gamma_{\delta} = \frac{1}{4}\omega_{0}^{2}S_{\delta\delta}(2\omega_{0}).
\]
Using the values from the preceding paragraph, ground state cooling requires approximately $\sqrt{S_{\delta\delta}(2\omega_{0})}<\num{1e-7}/\unit{\per\sqrt{Hz}}$ and $\sqrt{S_{\epsilon\epsilon}(\omega_{0})}<\qty{1e-19}{m\per\sqrt{Hz}}$ (corresponding to $T_{eff}<\qty{20}{mK}$). Regarding the former we note that relative fluctuations on the order of one part per billion can be reached in superconducting coils \cite{Britton2016}, while the latter requires further improvements to our vibration isolation system. Suspending the system from the top plate of the cryostat would allow us to add two more stages and achieve the required vibration suppression. 

\bibliography{ms}